\documentclass[a4paper,11pt]{article}
\pdfoutput=1
\usepackage{jheppub}
\usepackage[T1]{fontenc}
\usepackage{amsmath,amssymb,epsfig,txfonts}
\usepackage{xspace}
\usepackage{graphicx,epsfig}
\usepackage{subcaption}
\usepackage{verbatim}
\usepackage{textcomp}
\usepackage{booktabs}
\usepackage[normalem]{ulem}
\usepackage[section]{placeins}

\def\ttbar{\ensuremath{t\bar{t}}\;}
\def\etmiss{\ensuremath{E_{\mathrm{T}}^{\mathrm{miss}}}\;}
\def\ptmiss{\ensuremath{\vec p^{\mathrm{\ miss}}_\mathrm{T}}\;}
\def\ptl{\ensuremath{\vec p^{\mathrm{\ \ell}}_\mathrm{T}}\;}
\newcommand{\mttwo}{\ensuremath{m_{\mathrm{T2}}}\;}
\newcommand{\mttwoblj}{\ensuremath{m_{\mathrm{T2}_{blj}}}\;}
\newcommand{\amttwo}{\ensuremath{am_{\mathrm{T2}}}\;}

\newcommand{\mTlep}{\ensuremath{m_\mathrm{T}^{lep}}\;}

\newcommand{\btagged}{\ensuremath{b}-tagged\;}
\newcommand{\mbl}{\ensuremath{m_{b\ell}}\;}
\def\GeV{\ifmmode {\mathrm{\ Ge\kern -0.1em V}}\else
                   \textrm{Ge\kern -0.1em V}\fi}%
\makeatother
\begin{document}
\title{Single-top final states as a probe of top-flavoured dark matter models at
the LHC}

\author[a,b]{Monika Blanke,}
\author[c,d]{Priscilla Pani,}
\author[e]{Giacomo Polesello,}
\author[e,f]{Giulia Rovelli}

\affiliation[a]{Institute for Astroparticle Physics, Karlsruhe Institute of Technology,
  Hermann-von-Helmholtz-Platz~1,
  76344 Eggenstein-Leopoldshafen, Germany}
\affiliation[b]{Institute for Theoretical Particle Physics,  Karlsruhe Institute of Technology, Engesserstra{\ss}e 7,\\ 76131 Karlsruhe, Germany}
\affiliation[c]{Deutsches Elektronen-Synchrotron DESY, Notkestraße 85, 22607 Hamburg, Germany}
\affiliation[d]{Deutsches Elektronen-Synchrotron DESY, Platanenallee 6, 15738 Zeuthen, Germany}
\affiliation[e]{INFN, Sezione di Pavia, Via Bassi 6, 27100 Pavia, Italy}

\affiliation[f]{Universit\`a  degli Studi di Pavia, Dipartimento di Fisica, Via Bassi 6, 27100 Pavia, Italy}

\emailAdd{monika.blanke@kit.edu}
\emailAdd{priscilla.pani@cern.ch}
\emailAdd{giacomo.polesello@cern.ch}
\emailAdd{giulia.rovelli@cern.ch}

\preprint{TTP20-034, P3H-20-059}

\abstract{Models incorporating {flavoured dark matter} provide 
an elegant solution to the {dark} matter problem, evading {the} tight LHC and direct direction constraints on simple WIMP models. {In Dark Minimal Flavour Violation, a simple framework of flavoured dark matter with new sources of flavour violation,
the constraints} from thermal freeze-out, {direct detection experiments, and} flavour physics create well-defined benchmark scenarios for these models.
We study the LHC phenomenology of four such scenarios, focusing on final states where a single top quark is produced accompanied by no jets, one jet from the fragmentation of light quarks or a $b$-tagged jet.  For each of these signatures we
develop a realistic LHC analysis, and we show that the proposed analyses would
increase the parameter space coverage for the four benchmarks, compared 
to existing flavour-conserving LHC analyses. Finally we show the 
projected discovery potential of the {considered} signatures
for the full LHC statistics at 14~TeV, and for the High Luminosity LHC.}

\maketitle
\flushbottom

\section{Introduction}

Weakly interacting massive particles (WIMPs) are among the theoretically best-motivated candidates to explain the observed dark matter (DM) density in the universe  \cite{Arcadi:2017kky,Roszkowski:2017nbc}. However, the absence of signal in both direct detection experiments and at the Large Hadron Collider (LHC)  has put simple WIMP models under severe pressure, challenging the presence of a sufficiently large DM annihilation cross-section. 

A possible way out of this dilemma is offered by the introduction of a non-trivial flavour structure in the dark sector \cite{Kile:2011mn,Kamenik:2011nb,Batell:2011tc,Agrawal:2011ze,
Batell:2013zwa,Kile:2013ola,Lopez-Honorez:2013wla,Kumar:2013hfa,
Zhang:2012da}. In this scenario, dark matter transforms as a multiplet (usually triplet) under a flavour symmetry and couples non-universally to the different flavours of the Standard Model (SM). If the lightest dark flavour couples predominantly to the third quark generation, its interactions with the SM nuclei are reduced, thereby reconciling the thermal freeze-out condition with the experimental limits. Such models are usually referred to as top- or bottom-flavoured dark matter.

Most interesting from the phenomenological point of view are models which go beyond the assumption of Minimal Flavour Violation (MFV). In a series of papers \cite{Agrawal:2014aoa,Blanke:2017tnb,Blanke:2017fum} the framework of Dark Minimal Flavour Violation (DMFV) has been put forward.  In this class of models, the interaction between the dark matter flavour triplet and the SM quarks\footnote{A leptonic DMFV model has been suggested in \cite{Chen:2015jkt}.}, mediated by a coloured scalar $\phi$, constitutes the only new source of flavour violation, thus efficiently reducing the number of free parameters while at the same time conserving the rich phenomenology of a non-MFV model. The interplay of constraints from flavour physics, direct detection experiments and the thermal freeze-out condition then allows to place limits on the parameter space of the model, thereby creating benchmark scenarios to be targeted by future LHC searches.

At the LHC, DMFV models provide final state signatures involving the production 
of scalar mediators ($\phi$), each of which further decays 
into a quark and  a  fermionic dark matter particle ($\chi$).
The scalar mediators can be produced in pairs, leading to signatures
with two quarks and \etmiss, or singly produced accompanied 
by a dark matter particle, leading to final states with {one quark} and
\etmiss.\par

As $\phi$ has the same quantum numbers as a supersymmetric squark,
the final {states} for pair production, when both mediators
decay to experimentally indistinguishable quark flavours
{are} identical to {flavour-conserving} SUSY squark production, and 
limits on the parameter space of the DMFV models can be obtained 
by a simple recasting of the existing squark searches at the
LHC. This exercise was performed on the LHC Run 1 results in 
\cite{Agrawal:2014aoa,Blanke:2017tnb,Blanke:2017fum}, yielding
stringent limits on the parameter space of the model.

A specific feature of DMFV models is 
the flavour-violating signatures with a single quark or two 
quarks with different flavours. In the case of top-flavoured DM, this leads to
LHC signatures featuring a single top quark accompanied by 
two dark matter particles, and either zero additional jets 
or an additional light ($u$,$d$,$s$,$c$) or
$b$-jet produced in the resonant decay of the mediator, on which we concentrate in this paper.

The detailed analyses of top-flavoured dark matter coupling to left- \cite{Blanke:2017tnb} or right-handed quarks \cite{Blanke:2017fum} have identified the phenomenological sweet-spots in the parameter space of these models for which the constraints from flavour and dark matter experiments are satisfied. Based on these findings,  we define for this work four benchmark classes of
models for which we develop a search 
strategy at the LHC based on final states with a single top.

After defining the benchmarks, we study in detail their LHC 
phenomenology, addressing both the signatures from mediator pair production 
and the final states featuring a single top quark.
Based on this work, we define the LHC constraints on the benchmarks
by recasting recent results from the LHC
experiments based on $\sim$140~fb$^{-1}$ of data collected at $\sqrt{s}=13$~TeV.
For each of the single-top signatures, 
we then develop a realistic analysis strategy, leading to 
a comparison of their potential in constraining the
parameter space of the four benchmarks with the existing LHC bounds. 
We further predict the reach of the proposed signatures
for the projected full statistics of the LHC at 14 TeV, 300~fb$^{-1}$,
and for the High Luminosity LHC project, 3~ab$^{-1}$.


\section{DMFV models and definition of the benchmark scenarios}

In this section, we present the simplified models of top-flavoured dark matter introduced in \cite{Blanke:2017tnb,Blanke:2017fum} that we will use throughout this analysis. We start by recapitulating the basics of Dark Minimal Flavour Violation (DMFV), and then move on to briefly review the theoretical ingredients and phenomenological implications of the two models in which DM couples either to right- or left-handed top quarks.  The experimental constraints identified in \cite{Blanke:2017tnb,Blanke:2017fum} will guide us in deriving four viable benchmark scenarios for our analysis of LHC constraints and single-top signatures.

\subsection{The DMFV framework}

The original models of flavoured dark matter embedded the assumption of Minimal Flavour Violation (MFV) \cite{Buras:2000dm,DAmbrosio:2002vsn,Buras:2003jf,Chivukula:1987py,Hall:1990ac}: the SM Yukawa couplings were assumed to be the only source of flavour violation, and hence the flavour structure and phenomenology of those models was highly restricted. Subsequently, in order to allow for a richer flavour phenomenology, the concept of Dark Minimal Flavour Violation (DMFV) was introduced in \cite{Agrawal:2014aoa}. In this framework, the coupling matrix $\lambda$ between the SM quarks and the fermionic DM field $\chi$, transforming as triplet under a new flavour symmetry $U(3)_\chi$,  is assumed to be the only new source of flavour violation beyond the SM Yukawa couplings. As a consequence, the three dark flavours $\chi_i$ are nearly degenerate, with a mass splitting generated only by corrections of the form {$\eta\lambda^\dagger\lambda$. Here $\eta$ is a free parameter within the simplified model that would be determined by the choice of a UV completion.} 
Thanks to the DMFV flavour symmetry, the lightest state in the dark sector is stable {\cite{Batell:2011tc,Agrawal:2014aoa}} and is assumed to form the observed DM of the universe.

The interaction between $\chi$ and the SM quarks is mediated by a scalar $t$-channel mediator $\phi$ that carries QCD colour charge. Its electroweak quantum numbers determine whether the DM couples to right-handed up- or down-type quarks or to the left-handed quark doublets. We thus have two possible implementations of top-flavoured DM in the DMFV framework, with the lightest flavour of $\chi$ coupling either to the right- or the left-handed top quark. These two models, dubbed ``right-handed'' and ``left-handed'', are introduced next.

\subsection{Right-handed model}

In the right-handed model, the scalar mediator $\phi$ carries the same gauge quantum numbers as the right-handed up-type quarks. Thus the DM flavour triplet $\chi$ couples to the right-handed up-type quark triplet via the interaction term
\begin{equation}
\mathcal{L}_\text{RH} \supset - \lambda_{ij} \bar{u}_{Ri} \chi_j \phi + h.c.\,.
\end{equation}
Here $\lambda$ is a $3\times 3$ complex matrix that can be parametrised in terms of three diagonal couplings $D_{\lambda,ii}>0$, three flavour mixing angles $0\le\theta_{ij}\le \pi/4$ and three complex phases $0\le\delta _{ij}<2\pi$ as follows:
\begin{eqnarray}
 \lambda&=&U_{\lambda}D_{\lambda}\qquad\text{with}\label{eq:lambda}\\
D_{\lambda}&=&\text{diag}(D_{\lambda,11},D_{\lambda,22},D_{\lambda,33})\,, \\
U_{\lambda}
	   &=& \left( \begin{matrix} 1&0&0\\ 0&c_{23}&s_{23}e^{-i\delta_{23}} \\ 0&-s_{23}e^{i\delta_{23}}&c_{23} \end{matrix} \right)
	   \left( \begin{matrix} c_{13}&0&s_{13}e^{-i\delta_{13}} \\ 0&1&0 \\ -s_{13}e^{i\delta_{13}}&0&c_{13} \end{matrix} \right)
	   \left( \begin{matrix} c_{12}&s_{12}e^{-i\delta_{12}}&0 \\ -s_{12}e^{i\delta_{12}}&c_{12}&0 \\0&0&1 \end{matrix} \right)\label{eq:U}\,,         
\end{eqnarray}
where $c_{ij}=\cos \theta_{ij}$ and $s_{ij}=\sin \theta_{ij}$. Note that the mixing angles $\theta_{ij}$ have been constrained to be at most $\pi/4$ in order to ensure that the DM flavour $\chi_i$ couples predominantly to the $i$th quark generation. We hence name $\chi_3 \equiv \chi_t$ the top-flavoured state. As in \cite{Blanke:2017tnb}, we consider $\chi_t$ to be the lightest dark flavour, i.\,e.\ DM is top-flavoured\footnote{The possibility of charm-flavoured DM has been considered in \cite{Jubb:2017rhm}.}, which is favoured by the stringent upper limit on the WIMP-nucleus {scattering} cross-section from direct detection experiments.

\subsection{Left-handed model}

In the left-handed model, instead, the scalar mediator carries the same gauge quantum numbers as the left-handed quark doublets, i.\,e.\ it is introduced as an $SU(2)_L$ doublet {$\phi = (\phi_u,\phi_d)^T$}.
The coupling of the DM flavour triplet $\chi$ to the left-handed quarks then reads
\begin{eqnarray}
\mathcal{L}_\text{LH} &\supset& - \lambda_{ij} \bar{q}_{Li} \chi_j \phi + h.c.
\nonumber\\
&=& - \lambda_{ij} \bar{u}_{Li}\chi_j\phi_u - \tilde{\lambda}_{ij} \bar{d}_{Li}\chi_j\phi_d+h.c.\,.
\end{eqnarray}
Here, $\lambda$ is parametrised as in \eqref{eq:lambda}-\eqref{eq:U}, and 
\begin{equation}
\tilde\lambda = V_\text{CKM}^\dagger \lambda
\end{equation}
accounts for the CKM misalignment between left-handed up- and down-type quarks. Again we assume $\chi_3\equiv \chi_t$ to be the lightest dark flavour and refer to it as top-flavoured DM. It couples to the SM top quark via $\phi_u$ and to the SM bottom quark via $\phi_d$.

\subsection{Phenomenology and benchmark scenarios}

We now briefly summarise the phenomenology of both models, as analysed in detail  in \cite{Blanke:2017tnb,Blanke:2017fum}. The constraints on their parameter space will guide us in our choice of four benchmark scenarios, on which we will focus our subsequent study of LHC signatures.

In the right-handed (RH) model, the most relevant constraint from flavour physics is due to the data on neutral $D$ meson mixing, requiring the mixing angle $\theta_{12}$ to be small unless the first two generation couplings, $D_{\lambda,11}$ and $D_{\lambda,22}$ are nearly degenerate. The non-observation of WIMP DM in direct detection liquid xenon experiments requires a significant cancellation between tree-level and $Z$-penguin contributions to the DM-nucleon scattering cross-section, requiring in particular a non-zero mixing angle $\theta_{13}$. Concerning the early universe phenomenology, different scenarios can be distinguished, depending on the lifetime of the heavier dark flavours $\chi_{1,2}$. The thermal freeze-out condition depends on the number of dark flavours present at the time the dark sector decoupled from the visible sector. In \cite{Blanke:2017tnb} two benchmark cases were investigated: quasi-degenerate freeze-out (QDF), in which the decay of the heavy flavours takes place after the DM freeze-out, so that all three flavours contribute to the effective annihilation cross-section, and single flavour freeze-out (SFF), in which the heavier flavours have decayed into $\chi_t$ already in the equilibrium phase. These two cases are distinguished mainly by the mass splitting between the different flavours $\chi_i$, which in the DMFV framework is generated by the flavour non-universality in the couplings $D_{\lambda,ii}$: {While the QDF scenario requires a relative mass splitting of 1\% or less, in the SFF scenario a mass splitting of about 10\% is assumed.}\footnote{Note that even in the single-flavour freeze-out {scenario}, the decay of the heavy flavours is irrelevant for the purpose of the present paper as the visible decay products are too soft to be identified in the LHC searches discussed below. The distinctive LHC signatures of the heavier DM flavours decaying into the lightest state have been discussed in \cite{Kilic:2015vka}.}

Altogether, the combination of the various constraints leads us to the identification of two interesting benchmark scenarios for the right-handed model, summarised in table \ref{tab:benchmarks}. RH-SFF describes a typical set of parameters in the single-flavour freeze-out scenario, while RH-QDF picks a viable benchmark for the quasi-degenerate case.

\begin{table}
\centering{\begin{tabular}{|l|lll|}\hline
 & DM mass & couplings & mixing angles \\\hline\hline
{\bf RH-SFF} & $ m_\chi = 200\,\text{GeV} $ & $ D_{\lambda,11} = D_{\lambda,22}$ & $ \sin\theta_{13} = 0.25$ \\
& & $D_{\lambda,33} = D_{\lambda,11} +1.0 $ & $\theta_{12} = \theta_{23} = 0$\\\hline
{\bf RH-QDF} & $ m_\chi = 150\,\text{GeV} $ &  $ D_{\lambda,11} = D_{\lambda,22}$ & $ \sin\theta_{13} = 0.2$ \\
& & $D_{\lambda,33} = D_{\lambda,11} +0.2 $ & $ \theta_{12} = \theta_{23} = 0$\\\hline
{\bf LH-QDF1} & $ m_\chi = 150\,\text{GeV} $ & $ D_{\lambda,11} = D_{\lambda,22} $ & $ \sin\theta_{13} = 0.1$ \\
&& $D_{\lambda,33} = D_{\lambda,11} +0.1 $ & $ \theta_{12} = \theta_{23} = 0 $\\\hline
{\bf LH-QDF2} & $m_\chi = 450\,\text{GeV}$ & $ D_{\lambda,11} = D_{\lambda,22} $ & $\sin\theta_{13} = 0.2$ \\
&& $D_{\lambda,33} = D_{\lambda,11} +0.2 $ & $ \theta_{12} = \theta_{23} = 0 $\\\hline
\end{tabular}}
\caption{Definition of benchmark scenarios. In all cases the mediator mass $m_\phi$ and the coupling $D_{\lambda,11}$ are free parameters, while the complex phases $\delta_{ij}$ are set to zero.\label{tab:benchmarks}}
\end{table}

The left-handed (LH) model is more stringently constrained by flavour physics, due to its couplings to both up and down {quark} sectors. Consequently, the constraints from neutral kaon, $B_{d,s}$ and $D$ meson oscillations come into play. {The} unavoidable CKM misalignment between the couplings $\lambda$ and $\tilde\lambda$ {requires} the first two generations to be quasi-degenerate, {i.\,e.} $D_{\lambda,11}\simeq D_{\lambda,22}$. In this limit, the mixing angle $\theta_{12}$ becomes unphysical.  In addition, the data from $B_{d,s}$ meson mixing require the mixing angles $\theta_{13}$ and $\theta_{23}$ to be small unless the third generation, $\chi_t$, is quasi-degenerate with the first two. At the same time, as in the right-handed model, the required cancellation in the {WIMP-nucleus scattering} cross-section bounded by direct detection experiments demands a non-vanishing mixing angle $\theta_{13}$. The combination of direct detection and flavour physics constraints thus excludes the single-flavour freeze-out scenario in the left-handed model. We hence choose two benchmark scenarios within the quasi-degenerate freeze-out case, LH-QDF1 and LH-QDF2, that differ most notably in the DM mass.

\section{LHC phenomenology}
\label{sec:xsec}

At the LHC, the scalar mediator $\phi$ can be pair-produced by QCD interactions. It can also be pair- or singly produced through its interactions with the DM triplet $\chi$ and the SM quarks, parametrised by the matrix $\lambda$. In this section, we first perform a general study of the final states resulting from the pair production of mediators. After that, we concentrate 
on the final states featuring a single top quark, which can result either
from pair or single production of mediators. For all the considered 
final states, we {study the dependence of} the expected 
production rates {on the parameters of the model}
for the benchmark cases introduced in the previous section.

\subsection{Pair production of mediators}\label{sec:pair-prod}

\begin{figure}
\begin{center}
\includegraphics[width=0.4\textwidth]{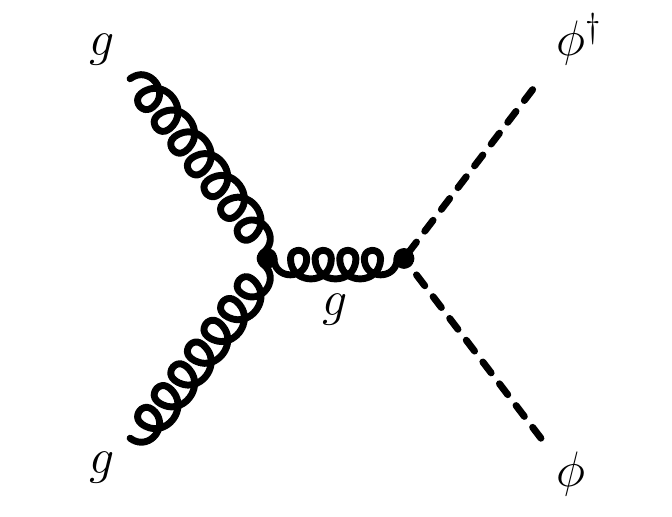}
\includegraphics[width=0.4\textwidth]{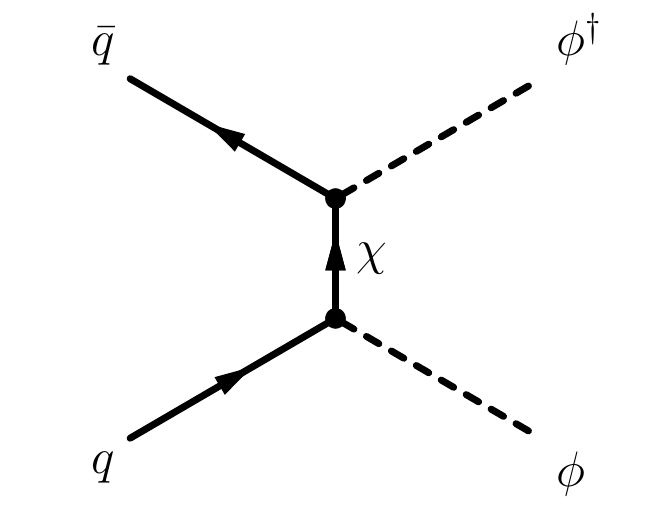}
\caption{Representative Feynman diagrams for the production of mediator pairs.}
\label{fig:feynchichi}
\end{center}
\end{figure}

The mediator $\phi$ can be pair-produced either through $s$-channel or through
$t$-channels processes. Representative diagrams for the two processes are 
shown in Figure~\ref{fig:feynchichi}.
The cross-section for $s$-channel production, a pure QCD process, is
independent of the parameters of the model, except for the mediator mass.
The cross-section for $t$-channel production depends instead on all 
parameters, in particular it has a quadratic dependence on
the product of the two relevant couplings.

\begin{figure}
\centering{\includegraphics[width=.48\textwidth]{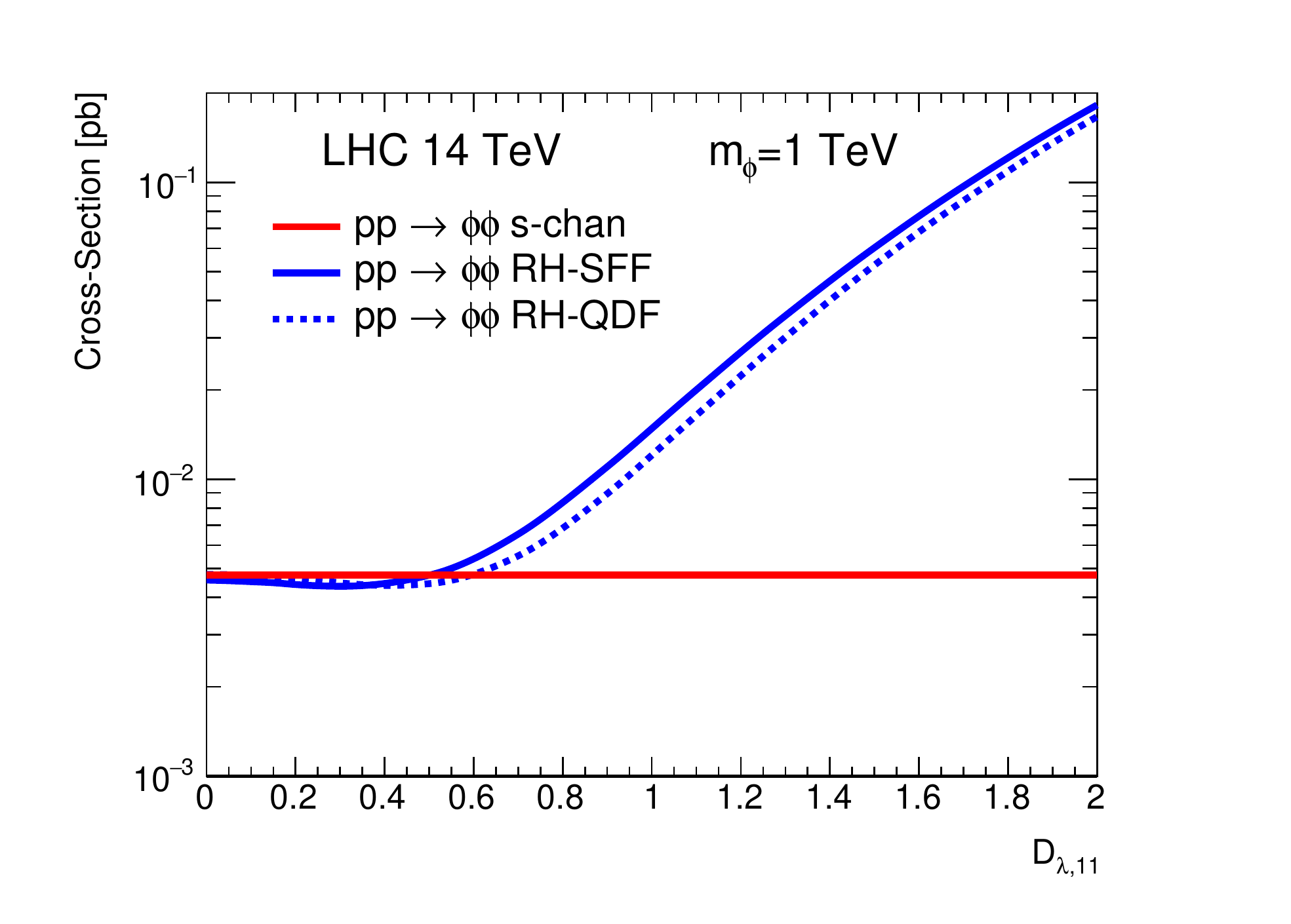}\quad
\includegraphics[width=.48\textwidth]{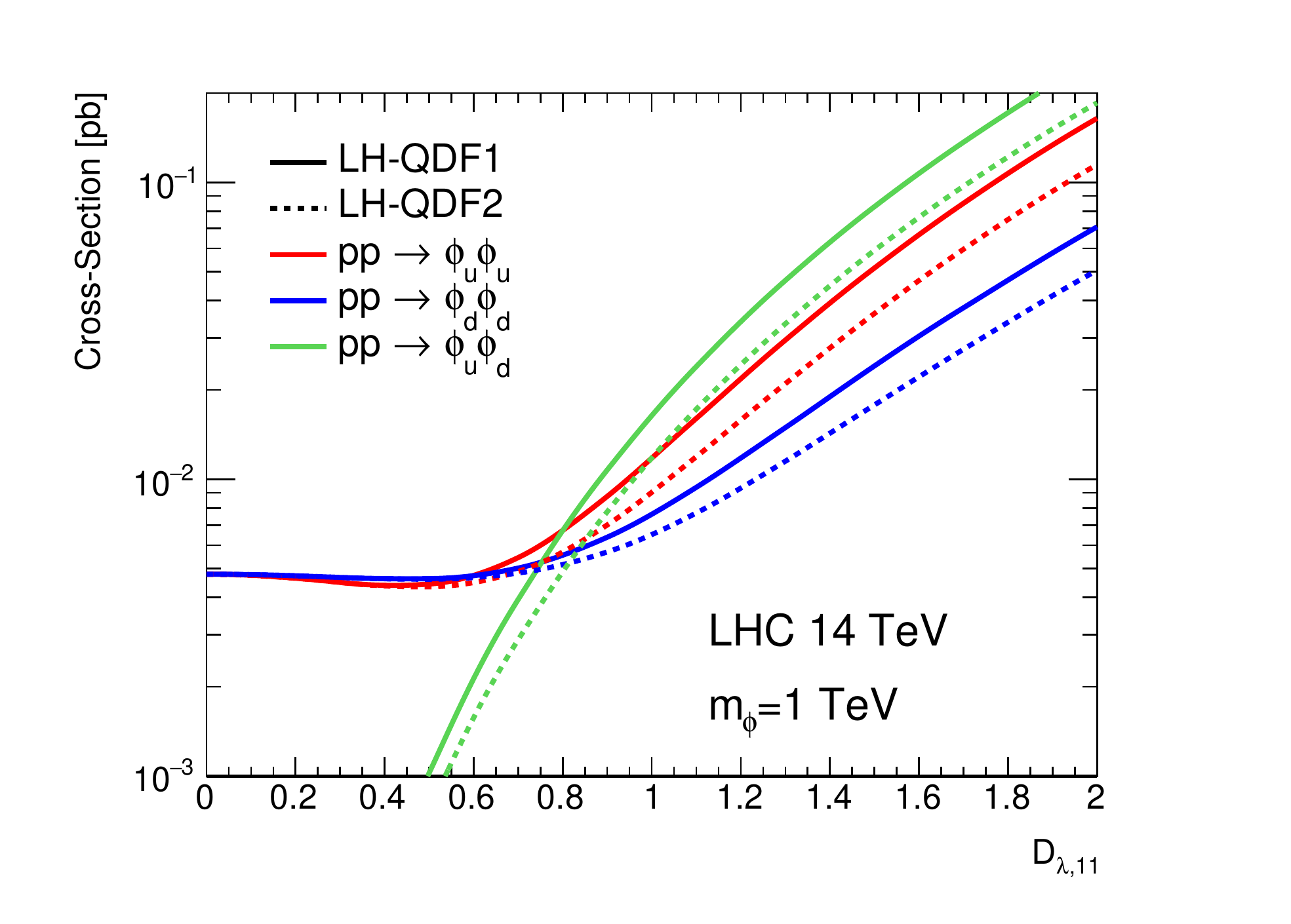}}
\caption{Cross-section for resonant production of two $\phi$ mediators as
a function of $D_{\lambda,11}$ at a 14~TeV LHC. 
The value of $m_\phi$ is fixed at 1 TeV.
Left: RH-SFF (solid line) and RH-QDF1 (dashed line) benchmarks. 
Right: Cross-section for three configurations: $\phi_u \phi_u$, $\phi_d \phi_d$, $\phi_u \phi_d$ for benchmarks LH-QDF1 (solid line) and LH-QDF2 (dashed line).\label{fig:xsec}}
\end{figure}

The total production cross-sections for the four benchmarks are shown 
in Figure~\ref{fig:xsec} as a function of the coupling $D_{\lambda,11}$.
Only one type of mediator $\phi$ is produced in the RH benchmarks. 
At low $D_{\lambda,11}$ the $t$-channel diagrams
slightly decrease the total cross-section through a negative interference with
the $s$-channel diagrams. With increasing $D_{\lambda,11}$ the $t$-channel becomes dominant,
{and it overtakes the $s$-channel for a value of $D_{\lambda,11}$ which depends
on $D_{\lambda,33}$},
due to the non-zero $\theta_{13}$ mixing angle.
In the case of the  LH benchmarks, three configurations for the pair production of mediators
are relevant: $\phi_u \phi_u$, $\phi_d \phi_d$, and $\phi_u \phi_d$. For the first
two, the dependence on $D_{\lambda,11}$ looks like the one discussed for 
the RH benchmarks, whereas $\phi_u \phi_d$ is a pure $t$-channel process and grows with $D_{\lambda,11}$ 
over the full considered $D_{\lambda,11}$ range. The cross-section of $\phi_u \phi_u$ for
LH-QDF1 is very similar to the one for RH-QDF, which has comparable parameters.
The cross-sections of all three processes are larger for LH-QDF1 than for LH-QDF2 at high $D_{\lambda,11}$
because of the larger value of the $\chi$ mass exchanged in the $t$-channel
for the latter benchmark.

The cross-sections need to be convoluted with the branching fractions of the
mediators into the different quark flavours to obtain the cross-sections
for the final states of interest.
The branching ratios (BR) are determined by the squares of the relative
values of the couplings $D_{\lambda,11}$ and $D_{\lambda,33}$.
In the case of equal couplings  $D_{\lambda,11}=D_{\lambda,22}=D_{\lambda,33}$,
the mediator would have 33\% BR into 3rd generation quarks and
66\% into quarks of the first two generations. 

Since the right-handed model features only one type of mediator, 
coupling to up-type quarks, we have three final states:
$jj+\etmiss$ (henceforth shortened to $jj$) , $t\bar{t}+\etmiss$
($t\bar{t}$),  and $tj+\etmiss$ ($tj$) where $j$ is a quark from the first
two generations.  An interesting configuration is realised for
$D_{\lambda,11}=D_{\lambda,22}=0.5\times D_{\lambda,33}$. This case
yields the most favourable situation for the 
$tj$ final state which has a 50\% BR, as compared to 25\% each for $jj$ and $t\bar{t}$.
The value of $D_{\lambda,11}$ for which this happens for each of the benchmarks
ultimately determines the phenomenology.

\begin{figure}
\centering{\includegraphics[width=.48\textwidth]{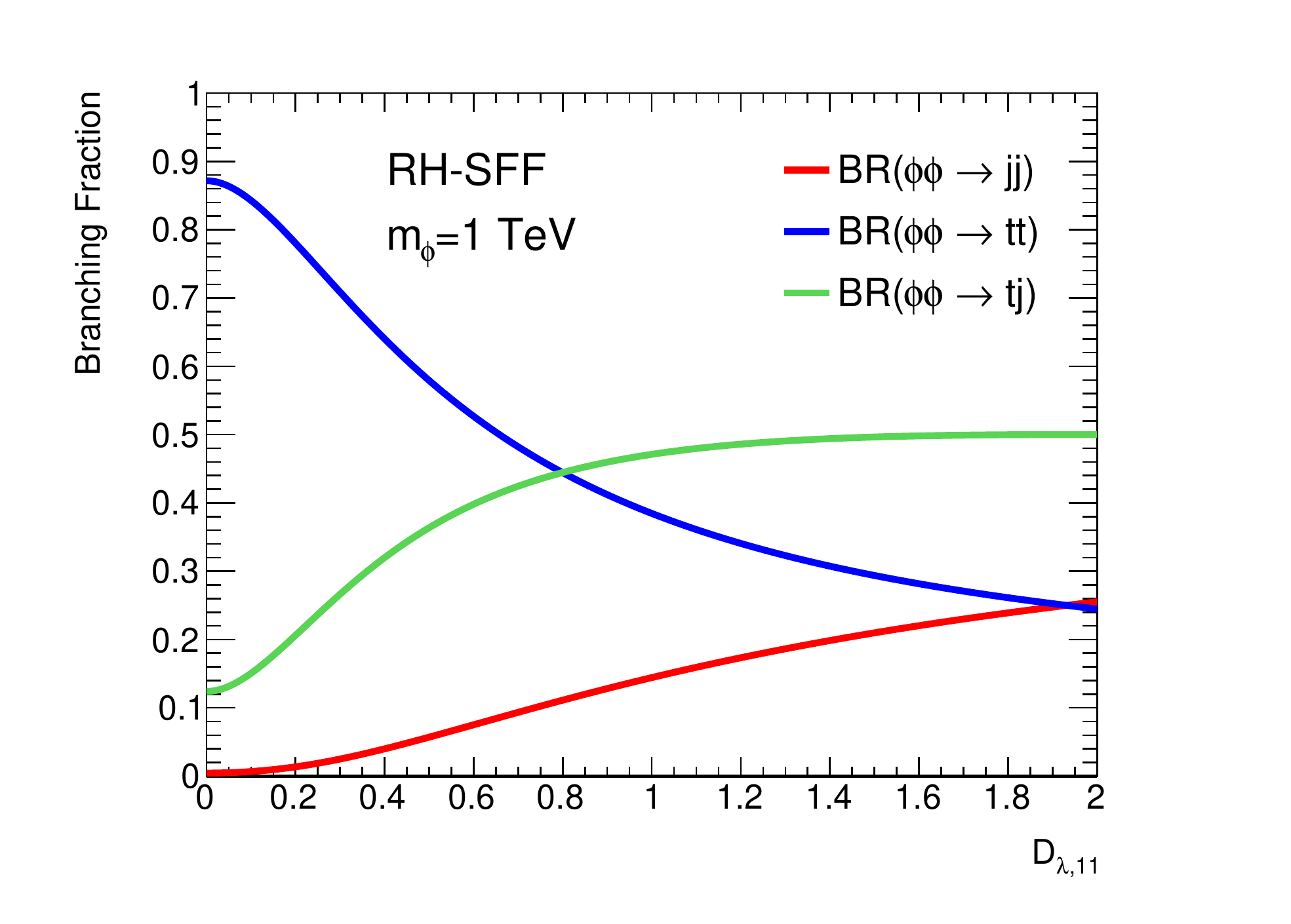}\quad
\includegraphics[width=.48\textwidth]{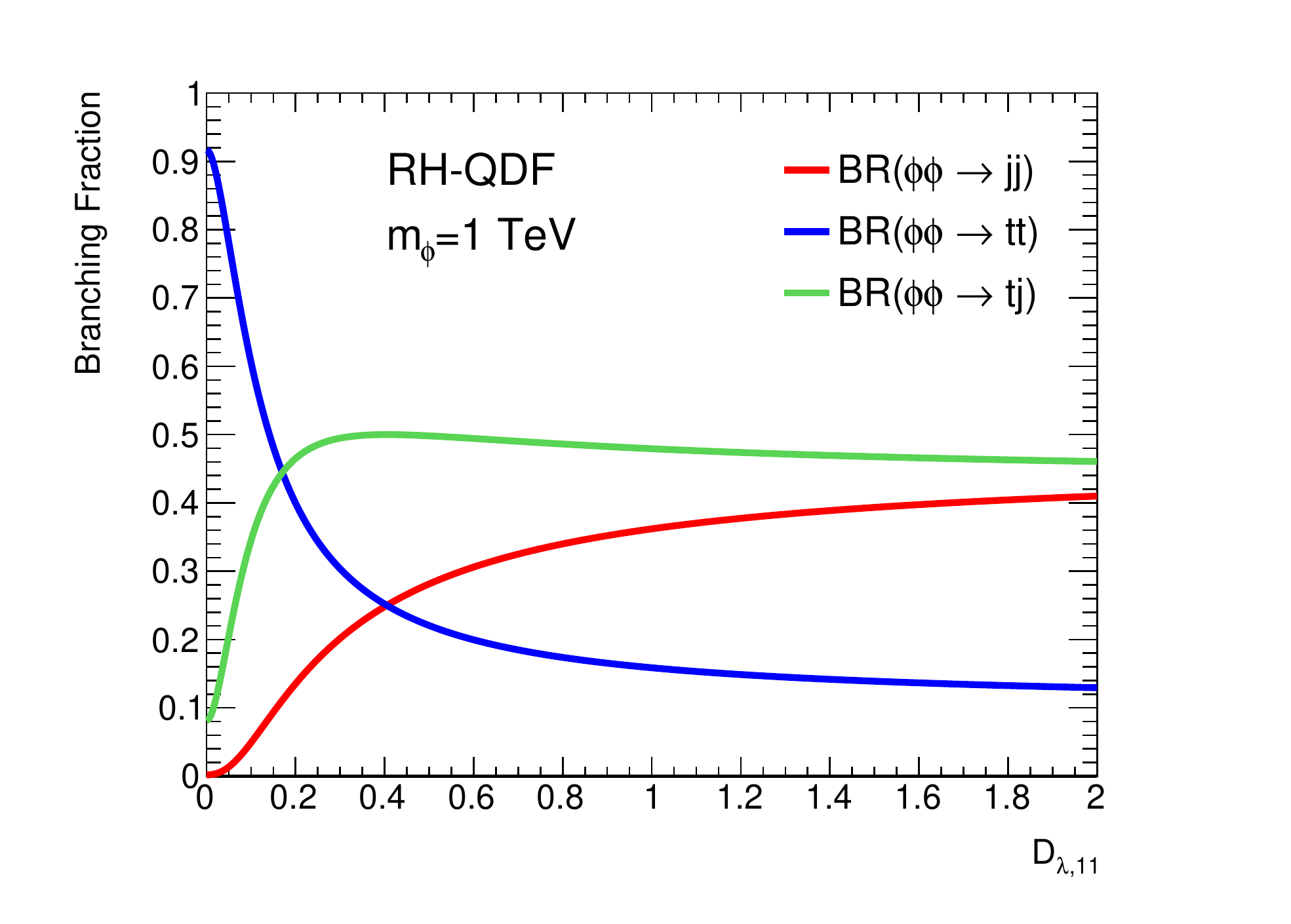}}
\caption{Branching fractions for three final states: $jj+\etmiss$, $tt+\etmiss$ and $tj+\etmiss$ 
as a function of the coupling $D_{\lambda,11}$ for the two RH benchmarks. Left: RH-SFF benchmark: right: RH-QDF benchmark.\label{fig:br:RH}}
\end{figure}

\begin{figure}
\centering{\includegraphics[width=.48\textwidth]{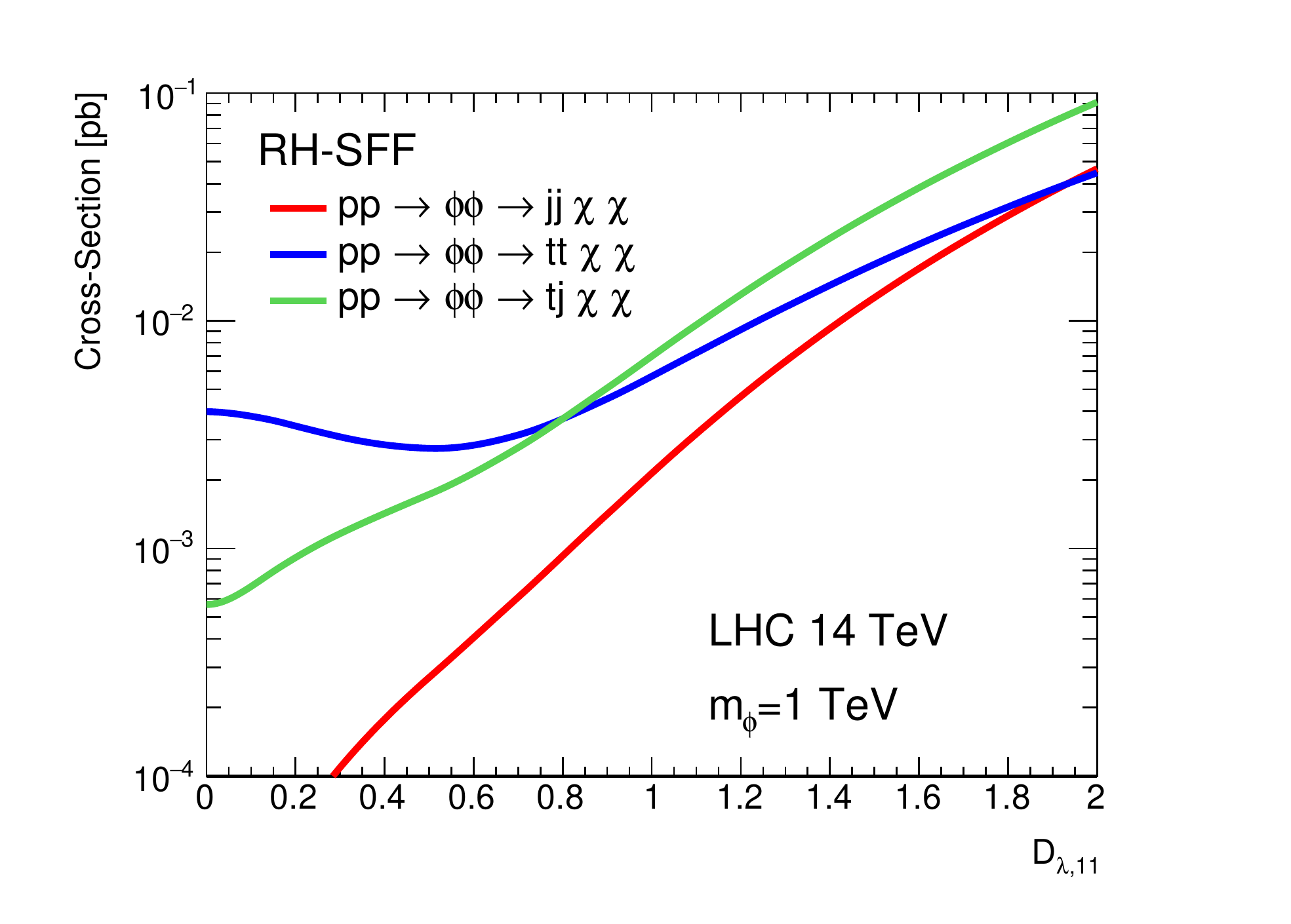}\quad
\includegraphics[width=.48\textwidth]{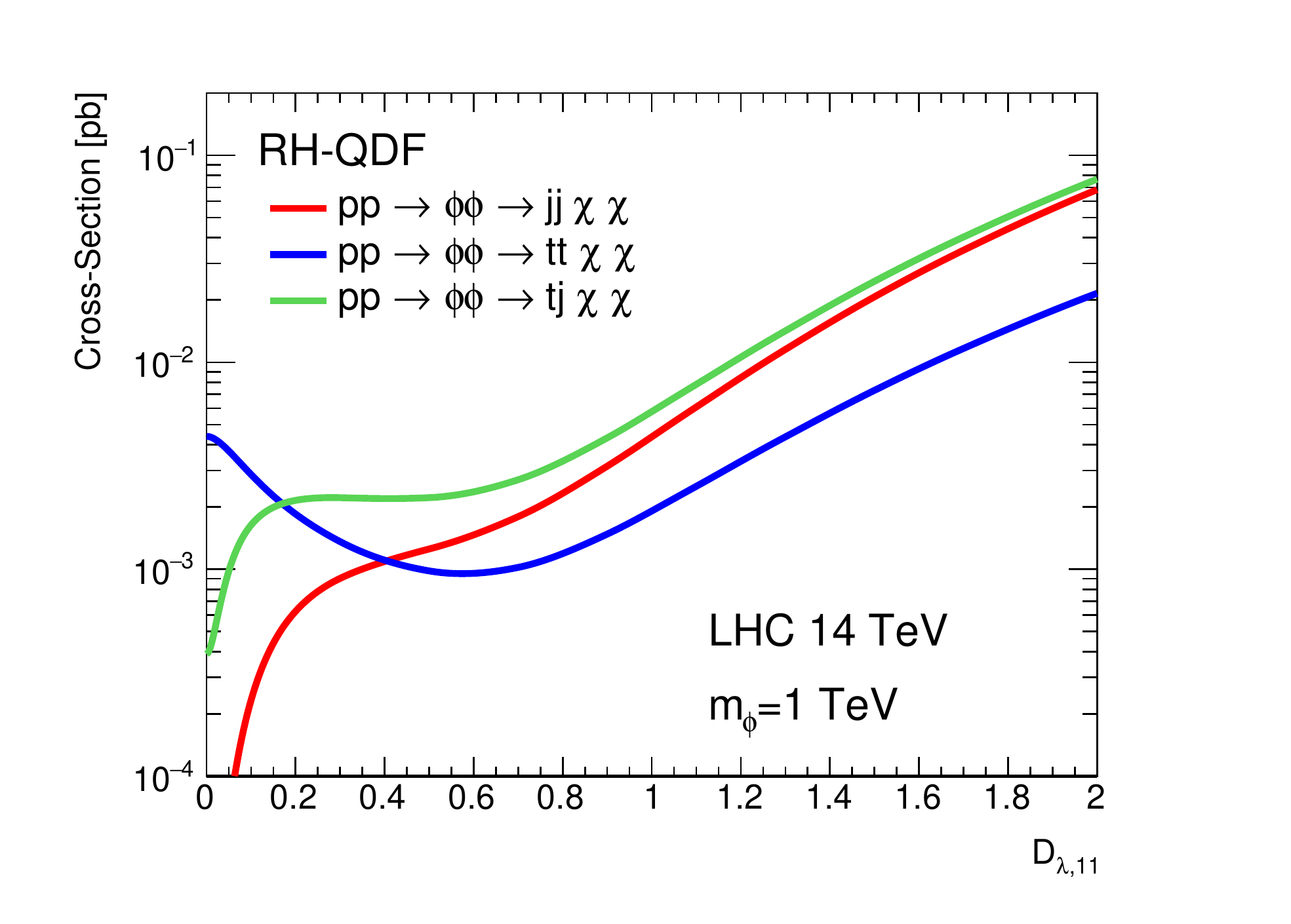}}
\caption{Production cross-section for the different final states resulting
from pair production of mediators at a 14 TeV LHC for the two RH 
benchmarks as a function of $D_{\lambda,11}$. Left: RH-SFF, right: RH-QDF.
The assumed mediator mass is 1~TeV.
\label{fig:xsbr:RH}}
\end{figure}

The branching fractions for the three final states in mediator pair
production are shown in Figure~\ref{fig:br:RH}.
In the RH-SFF benchmark (left), with $D_{\lambda,33}=D_{\lambda,11}+1$,  $t\bar{t}$ is dominant up to $D_{\lambda,11}=0.8$, where $tj$ takes over. 
For RH-QDF, with approximately balanced $D_{\lambda,11}$ and $D_{\lambda,33}$,
the switch between $t\bar{t}$ and $tj$ happens already for $D_{\lambda,11}=0.2$. 
Above this value, BR($tj$) is approximately flat at $\sim50\%$.
The non-zero value of $\theta_{13}$
is visible in the fact that the $t\bar{t}$ signature does not saturate 
the production for $D_{\lambda,11}\sim0$.
This behaviour translates directly into the cross-sections for the
three signatures, shown in Figure \ref{fig:xsbr:RH}.

\begin{figure}
\centering{\includegraphics[width=.48\textwidth]{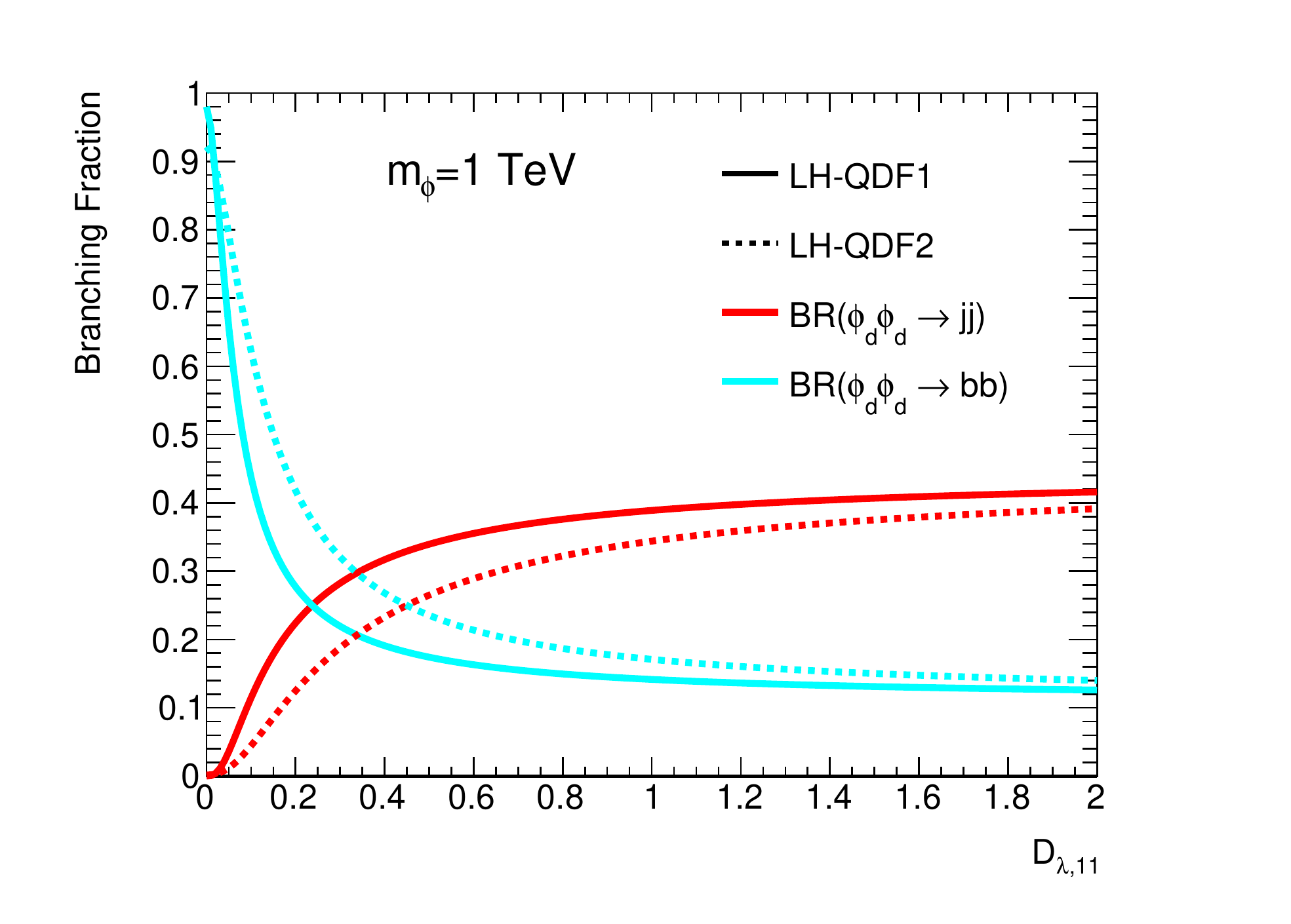}\quad
\includegraphics[width=.48\textwidth]{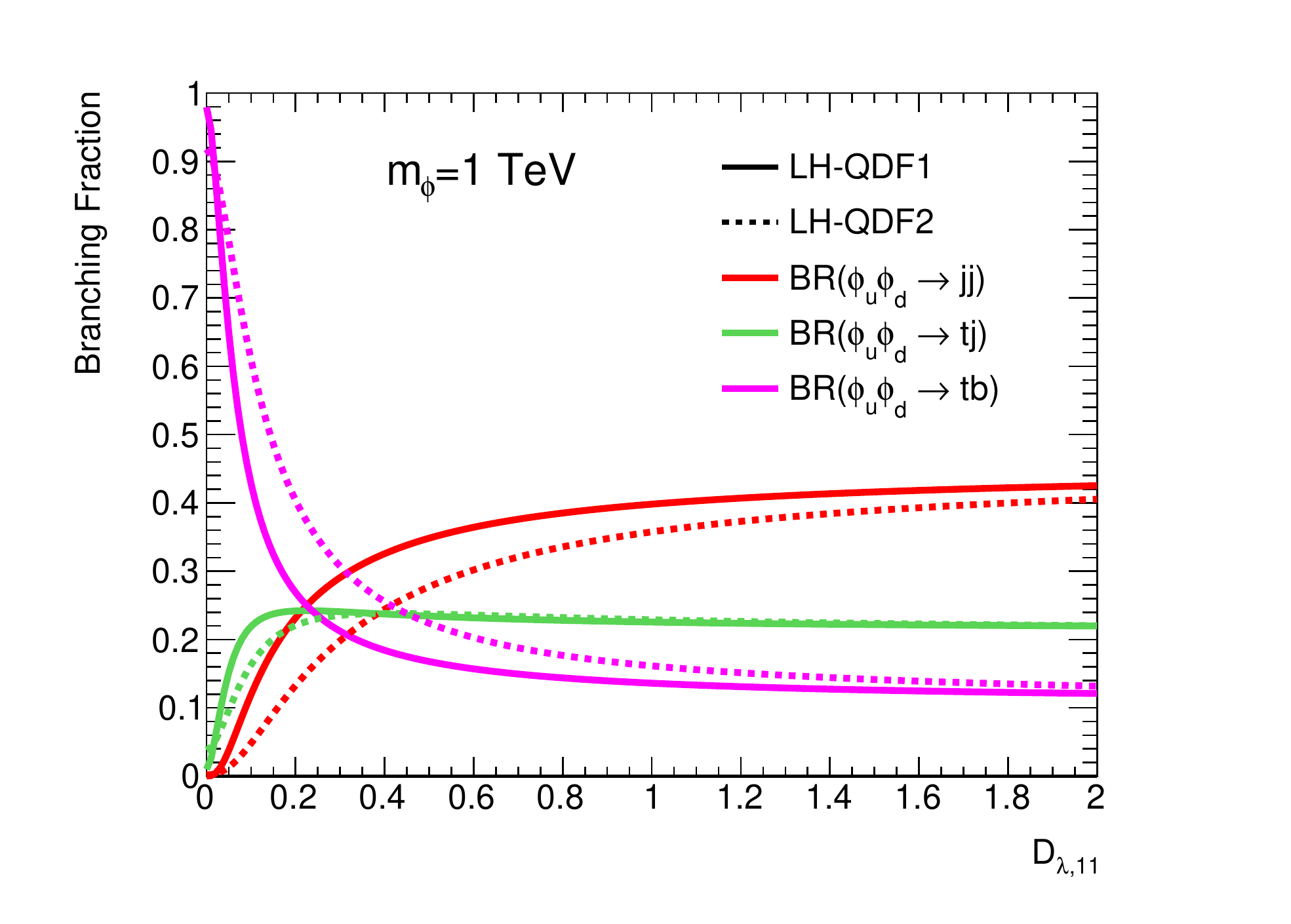}}
\caption{Branching fractions for the allowed final states respectively 
for $\phi_d \phi_d$ (left) and $\phi_u \phi_d$ (right) production
as a function of $D_{\lambda,11}$. The solid (dashed) lines are for 
LH-QDF1 (LH-QDF2). 
\label{fig:br:LH}}
\end{figure}

\begin{figure}
\centering{\includegraphics[width=.48\textwidth]{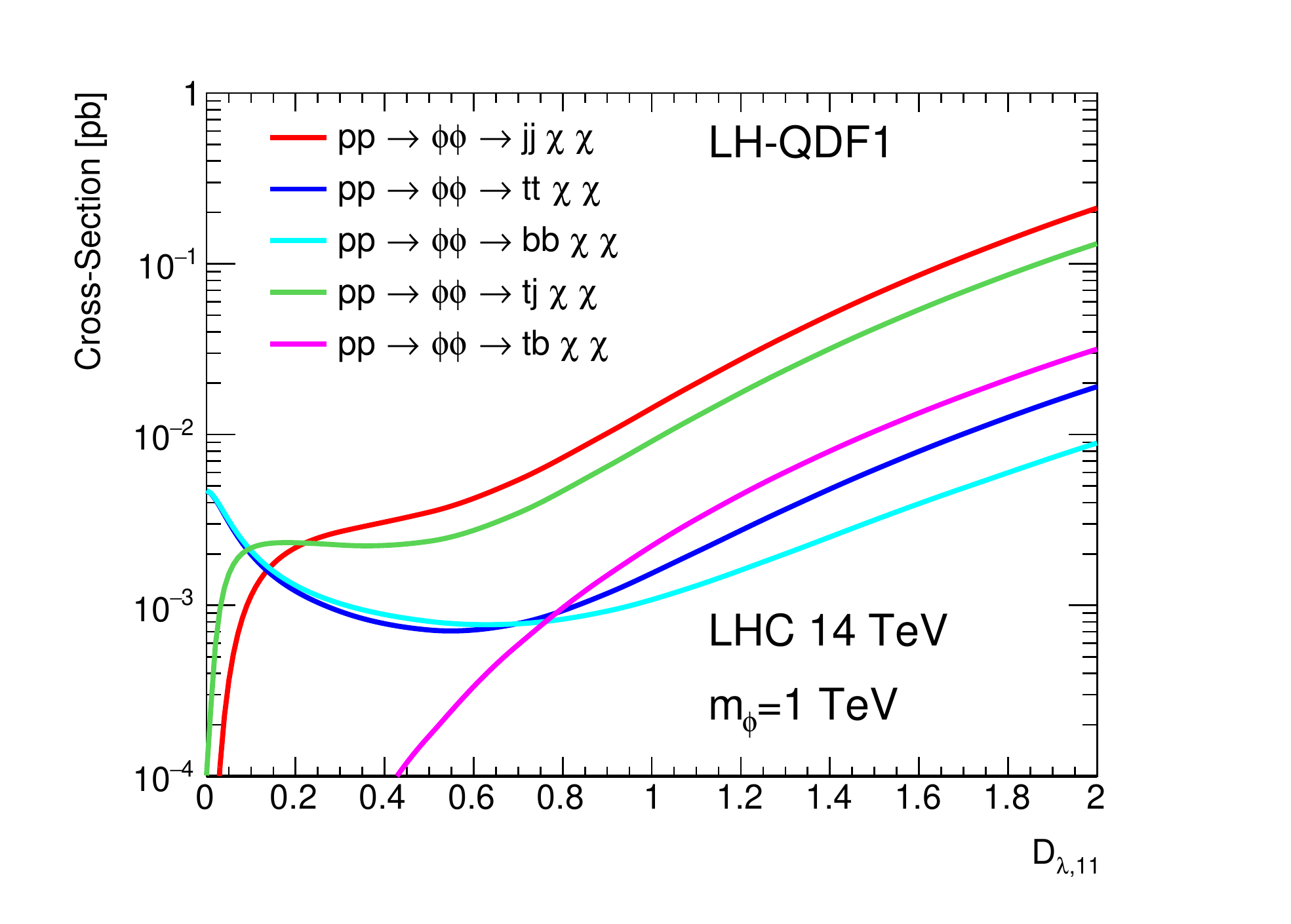}\quad
\includegraphics[width=.48\textwidth]{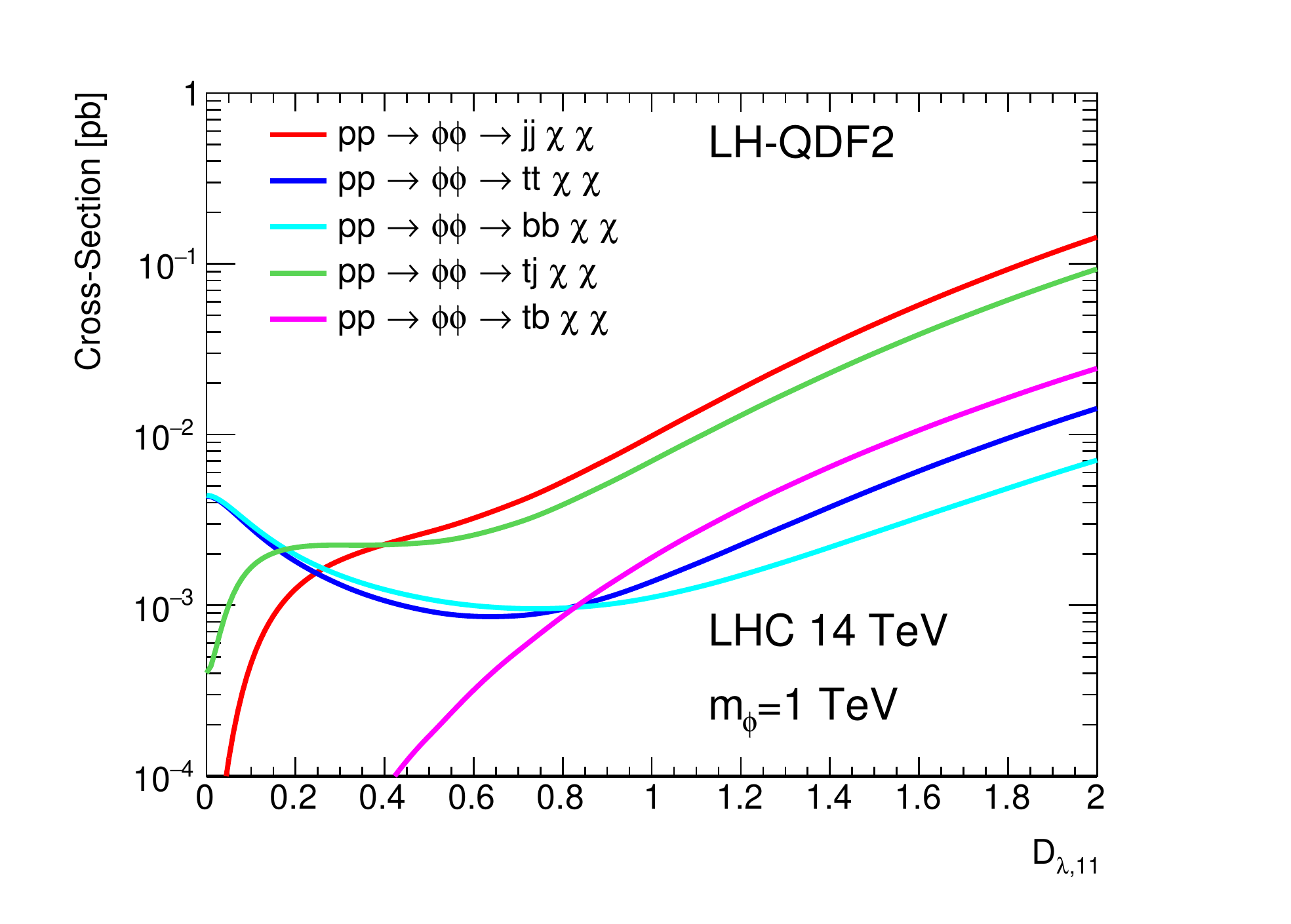}}
\caption{Production cross-section for the different final states resulting
from pair production of mediators at a 14 TeV LHC for the two LH 
benchmarks as a function of $D_{\lambda,11}$. Left: LH-QDF1, right: LH-QDF2.
The assumed mediator mass is 1~TeV.
\label{fig:xsbr:LH}}
\end{figure}

Due to the $SU(2)_L$ structure, the situation is more complex
in the case of the LH benchmarks. {In this case} we are dealing with three different production processes, 
$\phi_u \phi_u$, $\phi_d \phi_d$, $\phi_u \phi_d$, and each of them 
gives rise to different final states combinations. For the $\phi_u \phi_u$ process, the
allowed final states of interest are $jj$, $t\bar{t}$ and $tj$, and the BR pattern
is very similar to the one shown in the right panel of 
Figure~\ref{fig:br:RH}. For $\phi_d \phi_d$ and $\phi_u \phi_d$, different channels are
open, and the BRs are shown in Figure~\ref{fig:br:LH}.
The $jj$ final state is allowed in 
all three production processes, $tj$ in two, and the purely 
third generation signatures only in one. The final cross-sections for
the different signatures combining the three production processes are
shown in Figure \ref{fig:xsbr:LH}.
The coupling dependence of the five  relevant signatures is very similar
for the two benchmarks, with the cross-sections for LH-QDF2 somewhat
lower. The $jj$ final state is dominant starting from 
$D_{\lambda,11}=0.2-0.3$, with $tj$ being not far below. In the region where $jj$ and $t\bar{t}$ are of similar size, the $tj$ signature becomes dominant. For $D_{\lambda,11}$ close to zero,
$t\bar{t}$ and $b\bar{b}$ have the largest cross-sections.

\subsection{Single top final states}

Signatures including a single top quark can be produced in two 
ways in DMFV models: either through the on-shell production of a pair of 
mediators, one of which decays into a top quark and a dark matter
particle, as already introduced in Section \ref{sec:pair-prod}, or through the production of a single mediator, either 
accompanied by a top quark or by a dark matter particle.

We classify in the following the possible single top signatures
at the LHC. We neglect signatures where the top quark is produced 
in the decay of a mediator and the additional light or $b$-quark 
is produced in QCD radiation. {The latter} represent different final-state topologies
as the additional quark is not produced in the two-body decay 
of a heavy particle.
We identify the following three final states:
\begin{itemize}
\item $t$ + \etmiss;
\item $t$ + $\bar{q}$ + \etmiss, with $q$ = ($u$, $d$, $s$, $c$); 
\item $t$ + $\bar{b}$ + \etmiss.
\end{itemize}
In this section, we analyse the origin of each of the signatures above in 
the right-handed up model and in the left-handed model. In section \ref{sec:ana}, we will 
then develop an LHC analysis strategy for each of the signatures 
and evaluate the reach in parameter space for the full LHC 
statistics and for the HL-LHC run.

\subsubsection*{\boldmath $t$ + \etmiss}

\begin{figure}
\begin{center}
\begin{subfigure}{.305\textwidth}\centering
\includegraphics[width=\textwidth]{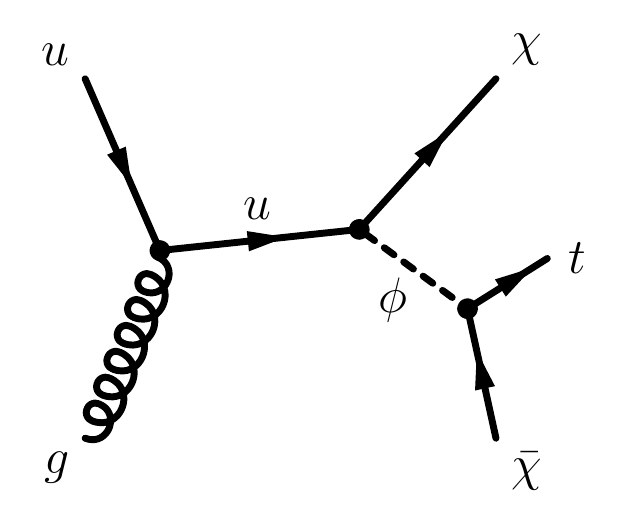}
\caption{}
\end{subfigure}
\begin{subfigure}{.30\textwidth}\centering
\includegraphics[width=\textwidth]{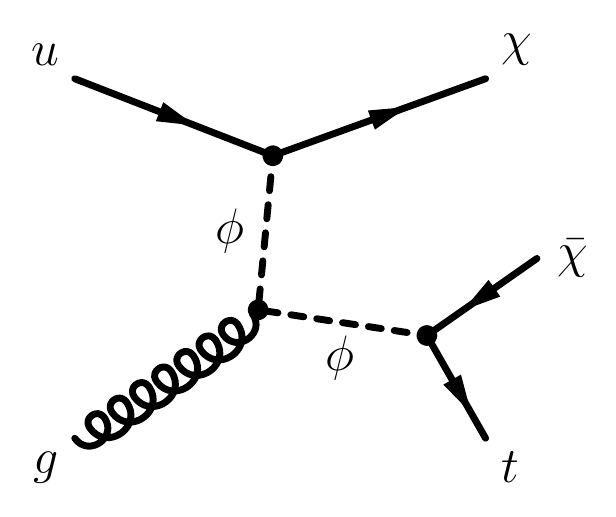}
\caption{}
\end{subfigure}
\begin{subfigure}{.30\textwidth}\centering
\includegraphics[width=\textwidth]{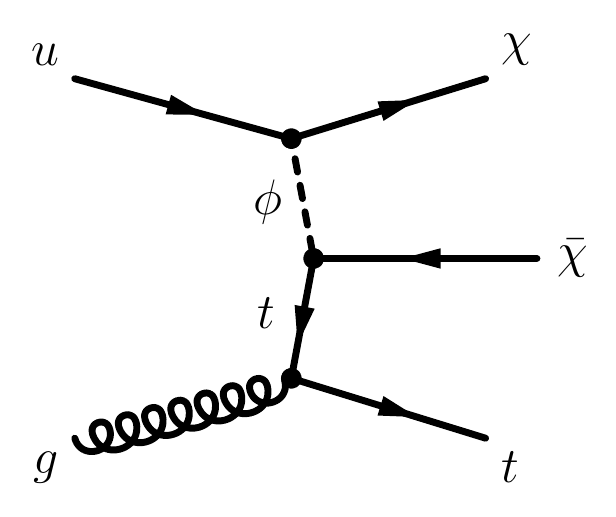}
\caption{}
\end{subfigure}
\caption{Representative Feynman diagrams for the production of two dark matter particles $\chi$ in association with a single top quark.}
\label{fig:feynmonotop}
\end{center}
\end{figure}

Representative Feynman diagrams
leading to the $t$ + \etmiss final state are shown in
Figure~\ref{fig:feynmonotop}.
The initial state is always a quark and a gluon.
For the first two diagrams on the left, (a) and (b), the top quark is produced
in the decay of an on-shell mediator, whereas diagram (c) features
a mediator exchanged in the $t$-channel.

The signature consisting in a single top quark recoiling
against two dark matter particles is frequently dubbed mono-top signature \cite{Andrea:2011ws,Alvarez:2013jqa}, 
and is similar to the one exploited in the ATLAS and CMS searches documented 
in \cite{Aaboud:2018zpr,Sirunyan:2019gfm}.
The signals considered in these studies, however, have a final state kinematics
which is different from the one of the  production processes shown in 
Figure~\ref{fig:feynmonotop}. Therefore, rather than attempting a recasting, 
we will develop a dedicated analysis strategy optimised for 
our benchmarks and based on the semileptonic decay of the top quark.

\subsubsection*{\boldmath $t$ + $\bar{q}$ + \etmiss}

\begin{figure}
\begin{center}
\begin{subfigure}{.30\textwidth}\centering
\includegraphics[width=\textwidth]{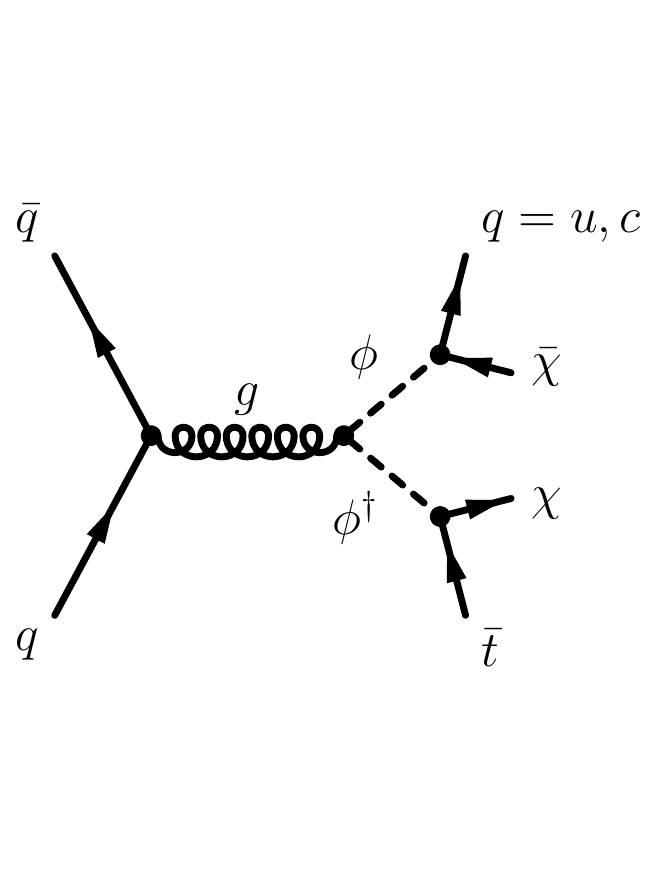}
\vspace{-50pt}
\caption{}
\end{subfigure}
\begin{subfigure}{.35\textwidth}\centering
\includegraphics[width=\textwidth]{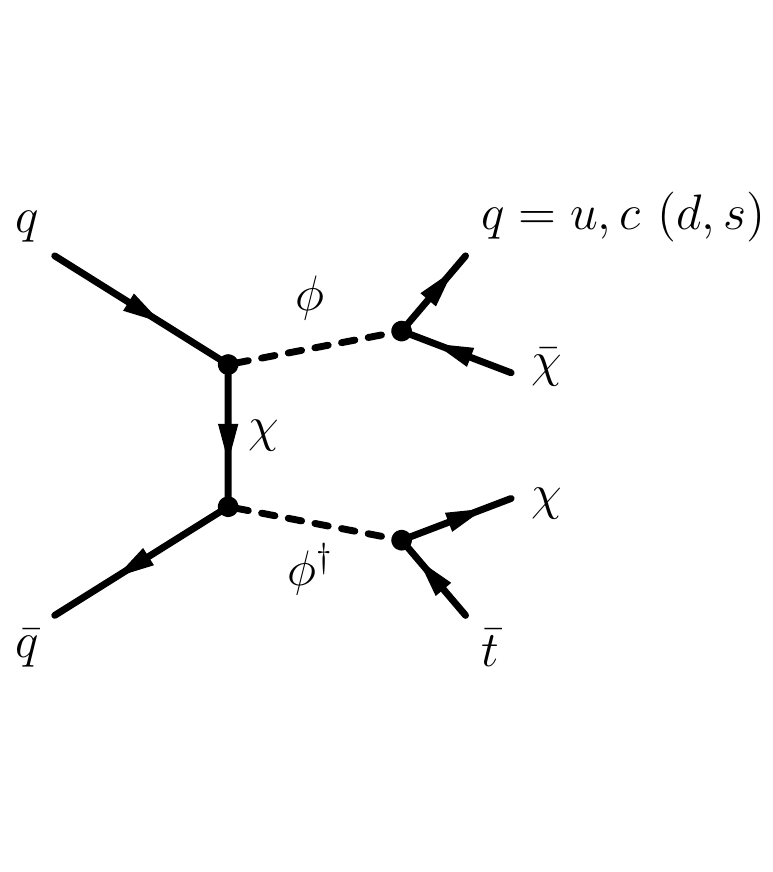}
\vspace{-50pt}
\caption{}
\end{subfigure}
\begin{subfigure}{.32\textwidth}\centering
\includegraphics[width=\textwidth]{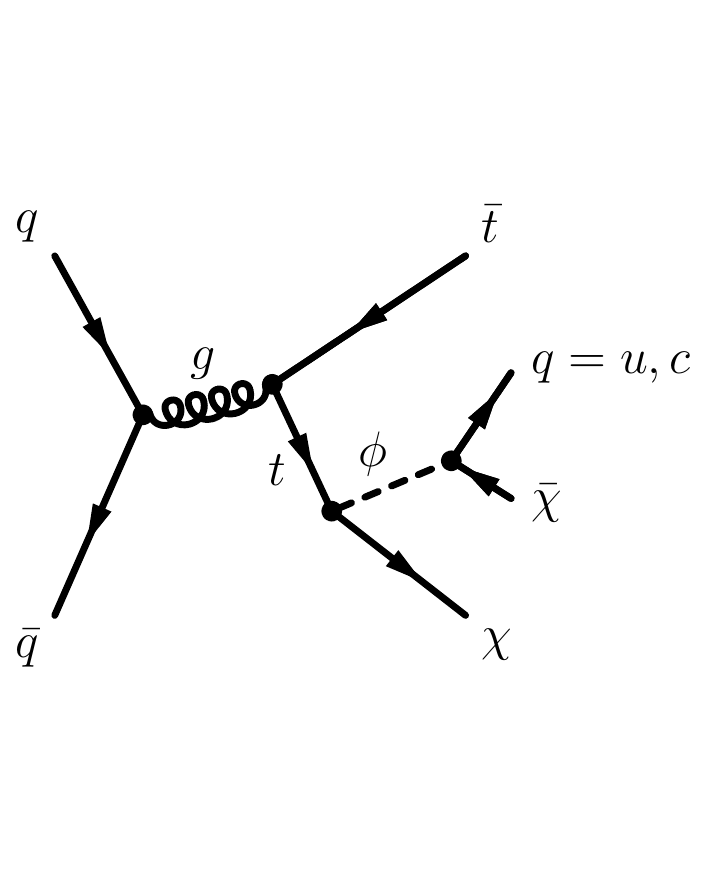}
\vspace{-50pt}
\caption{}
\end{subfigure}
\caption{Representative Feynman diagrams for the production of DM in association with a light jet and a  top quark.}
\label{fig:feyntj}
\end{center}
\end{figure}

Representative Feynman diagrams 
leading to the final state $t$ + $\bar{q}$ + \etmiss are shown in 
Figure~\ref{fig:feyntj}. 
The diagram in subfigure (a) depicts the pure QCD production
of two on-shell mediators and is equivalent to the 
non-MFV SUSY production of two different-flavour quarks \cite{Hurth:2009ke,Blanke:2013uia}.
Process (b) features the pair production
of two mediators with a dark matter particle exchanged in the $t$-channel.
{The final states with a $d$ or $s$ quark arise only in the LH model.}
For the diagram of  subfigure (c), the mediator is radiated from 
a $t$-quark leg in \ttbar production. 
The experimental reach for the process in subfigure (a) was studied in \cite{Chakraborty:2018rpn} 
in the framework  of a non-MFV simplified SUSY model, but it has not 
yet been the subject of an analysis by the LHC experiments.

Building on the analysis of \cite{Chakraborty:2018rpn}, we present
a re-optimisation of the selections based on a detailed simulation
of all the processes contributing to the addressed final state.

\subsubsection*{\boldmath $t$ + $\bar{b}$ + \etmiss}

\begin{figure}
\begin{center}
\begin{subfigure}{.40\textwidth}\centering
\includegraphics[width=\textwidth]{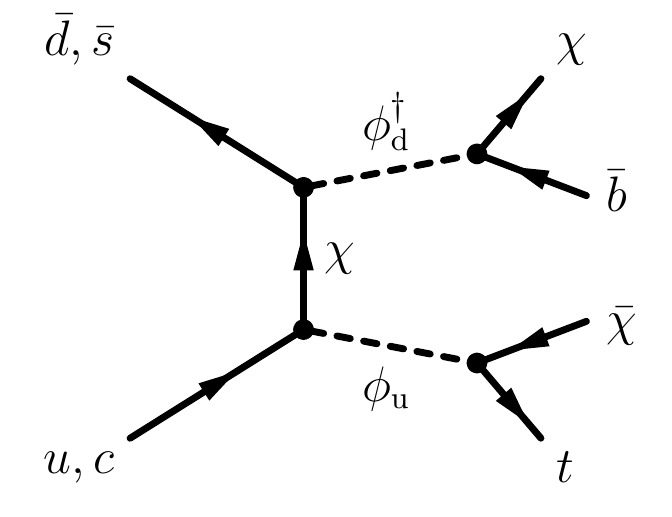}
\end{subfigure}
\caption{Feynman diagram for the production of DM in association with a  top and a $b$-quark.}
\label{fig:feyntb}
\end{center}
\end{figure}

This final state is only produced in the left-handed version of the 
model.  A representative Feynman diagram
leading to the final state $t$ + $\bar{b}$ + \etmiss is shown in 
Figure~\ref{fig:feyntb}.
Since the $b$-jet is identifiable experimentally, no $s$-channel production 
mechanism is available, differently from the $t$ + $\bar{q}$ + \etmiss
channel.

The $t$ + $\bar{b}$ + \etmiss signature is  specific to this model and does not arise from pair-production of SUSY squarks. It was studied
by the ATLAS experiment \cite{Aaboud:2017wqg}, however, in the framework of
the searches for cascade decays of the sbottom quark, targeting
SUSY models with compressed mass spectra. The addressed 
kinematic is very different from the one for 
the DMFV $t$ + $\bar{b}$ + \etmiss final state. We will therefore develop
an independent analysis strategy.\vspace{2ex}

\begin{figure}
\centering{\includegraphics[width=.48\textwidth]{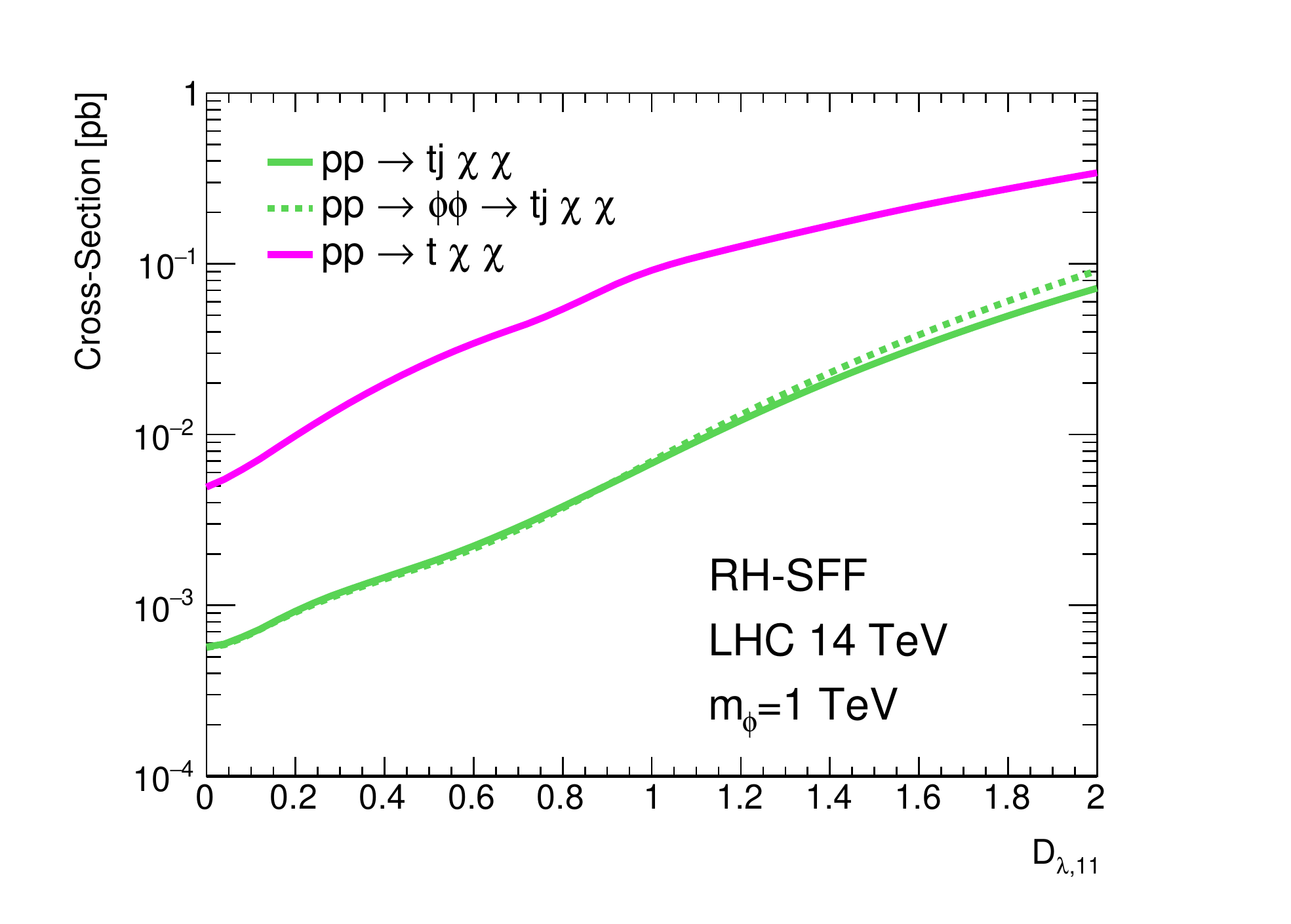}\quad
	\includegraphics[width=.48\textwidth]{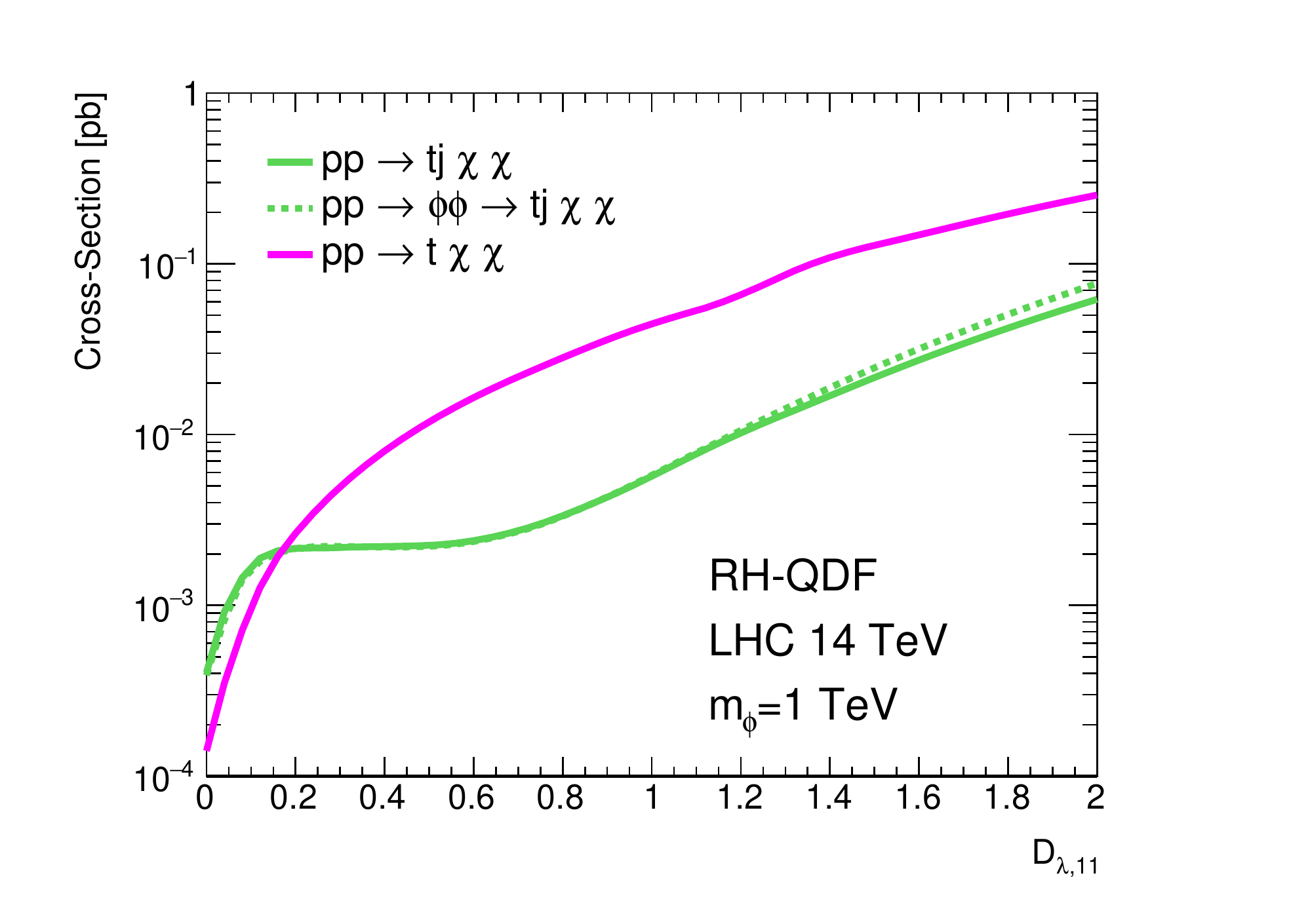}
\includegraphics[width=.48\textwidth]{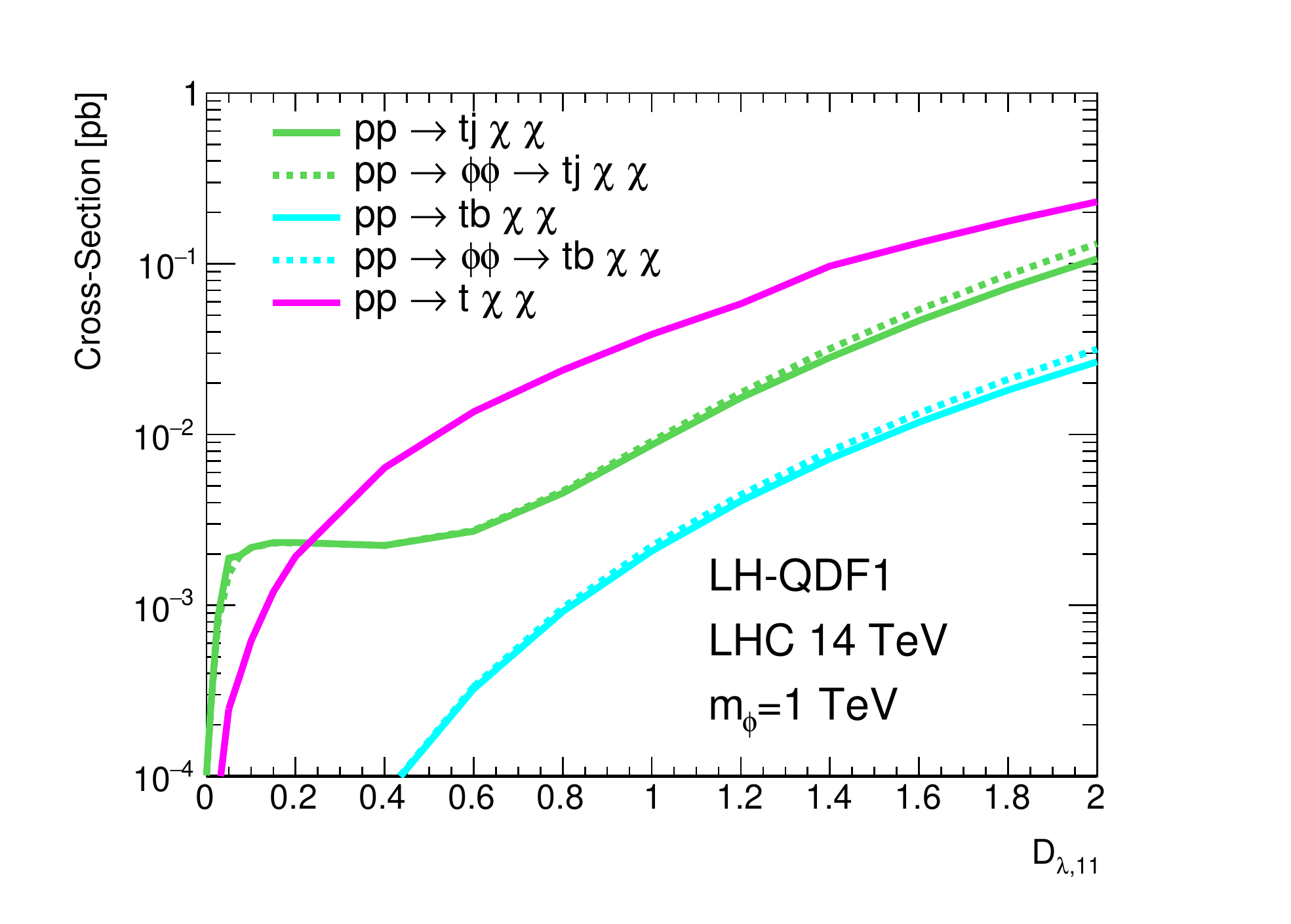}\quad
\includegraphics[width=.48\textwidth]{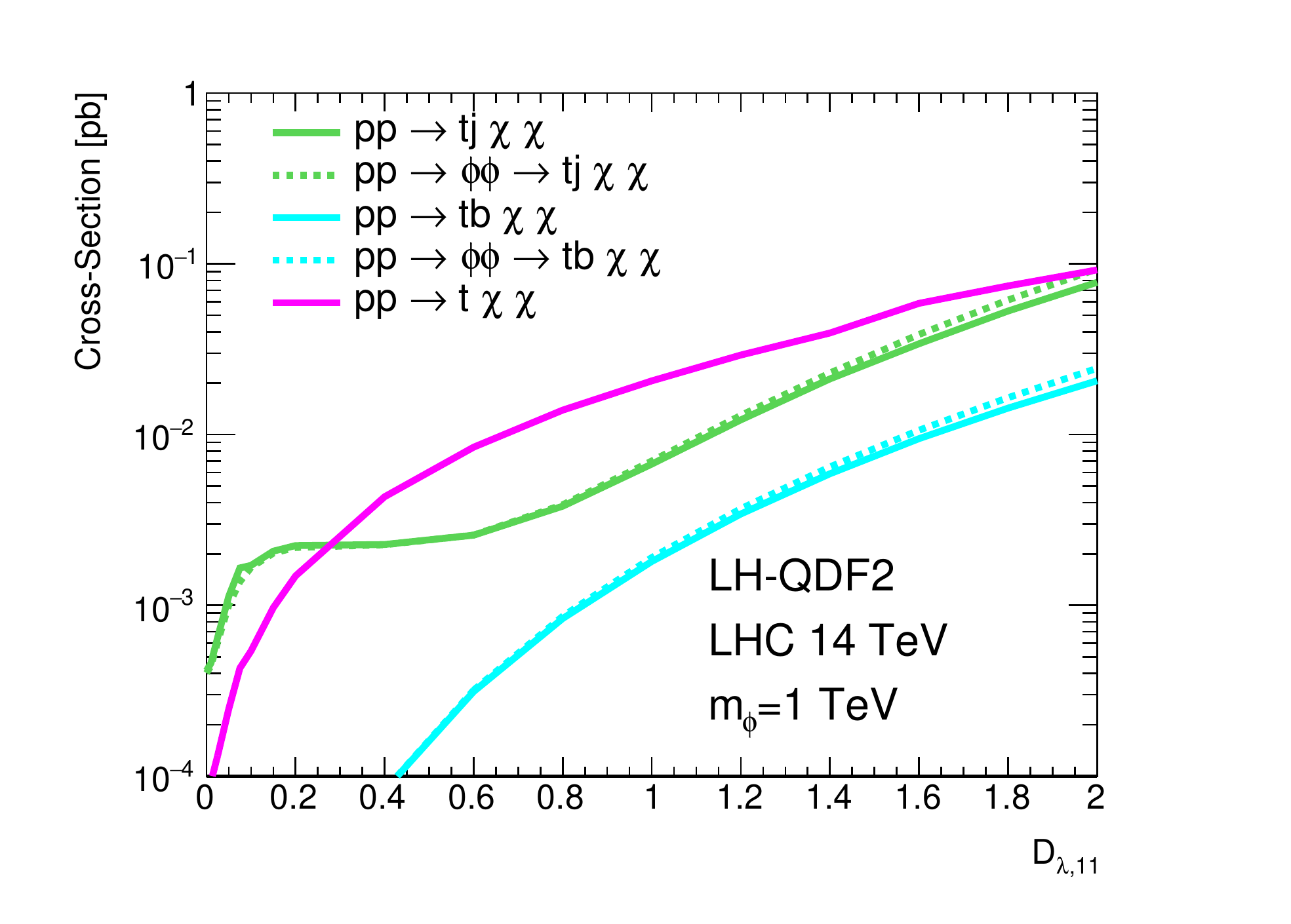}}
\caption{Production cross-section for the single top signatures in 14 TeV proton-proton collisions for the four benchmarks, as a function of $D_{\lambda,11}$. For the $tj$ and $tb$ final states both the value for the full cross-section and the one  through resonant production of mediator pairs is shown. The value of $m_\Phi$ is fixed at 1 TeV.\label{fig:xsec:xssingletop}
}
\end{figure}

The cross-sections for the three processes discussed above are shown
in Figure~\ref{fig:xsec:xssingletop} for the four benchmarks.
{In what follows}, we will use the shorthand {notation} \emph{monotop}, $tj$ and $tb$, respectively, for the three signatures.

For $tj$ and $tb$, the full production cross-section, shown as a solid
line, is compared to the one from mediator pair production, shown 
as a dashed line. Both cross-sections are dominated by 
the doubly-resonant component, and the discussion in 
section \ref{sec:pair-prod} applies.

The monotop signature is dominant over the whole {parameter} range considered for {the RH-SFF benchmark}. 
The three remaining benchmarks have a similar pattern, with monotop
dominating $tj$ down to $D_{\lambda,11}=0.1-0.2$. The monotop 
cross-sections are similar for RH-QDF and LH-QDF1, while they are
a factor 2-3 lower for LH-QDF2, due to the {much larger mass of $\chi$.}

Last but not least, the $tb$ signature in the LH model has a significantly smaller cross-section than both monotop and $tj$ over the full range of parameters considered.

\section{{Recast} of LHC limits for mediator pair production}

We consider the four benchmarks described in the previous section
and we explore the existing LHC bounds on these two models from
the on-shell productions of two coloured
mediators $\phi$ which in turn decay into a quark and a
dark matter particle {$\chi$}, which is the same final state
studied by SUSY squark searches.




The recast of  published searches for the production of two squarks 
relies on the assumption that the selection efficiency of each of the considered
analyses is the same for our model and for the simplified
SUSY ones used in the experimental analysis,
assuming the mass values such that $m_{\tilde{q}}=m_{\phi}$ and
$m_{\tilde{\chi}^0_1}=m_{\chi}$.

In all four benchmarks, for each mass of the mediator $\phi$ and  
each $D_{\lambda,11}$ value the cross-section for the process $$
pp \rightarrow \phi\phi^{\dagger} 
$$
is calculated for the relevant LHC centre-of-mass energy, as well 
the branching fractions of $\phi \rightarrow q \chi_q$, where q 
runs over all the quark flavours relevant for each benchmark,
as discussed in section \ref{sec:pair-prod}.
Based on the production cross-sections and branching fractions,
the total cross-section for each of the considered three final states is
calculated.


{The published LHC SUSY searches are in the framework of MFV models.
Therefore, the experimental SUSY results relevant for this study 
are the searches for pair productions of squarks of the 
first two generations, of stop squarks, and of sbottom squarks, addressing 
the final states $jj$, $t\bar{t}$ and $b\bar{b}$ of section~\ref{sec:pair-prod}.}
At the time of writing, only a limited set of relevant analyses based
on the full LHC Run 2 statistics of $\sim$140~fb$^{-1}$ have been
published \cite{Sirunyan:2019ctn, Sirunyan:2019glc, Sirunyan:2019xwh,
Aad:2020sgw}. A necessary condition for the present study 
is the availability in tabular form of the limits on the SUSY cross-section 
over a broad range of squark masses, extending down to $\sim$500 GeV
for the values of {$m_{\tilde q}$} assumed by our benchmark
models. The CMS study published in \cite{Sirunyan:2019ctn} 
presents a search for strongly produced SUSY in final states 
with multiple hadronic jets and \etmiss based 
on a statistics of 137~fb$^{-1}$. The direct production of 
squarks in simplified models resulting in all of the three final 
states of interest, $jj$+\etmiss, $\ttbar$+\etmiss and $b\bar{b}$+\etmiss ,
is explicitly addressed  in the paper, and tables of cross-section limits 
on very fine grids in ($m_{\tilde q}$,$m_{\tilde\chi^0_1}$) are provided 
for masses of squarks of the first two generations between 500 and 2000~GeV, going down to even lower masses
for stop and sbottom. We base our recasting study on this work.
For stop final states, a dedicated ATLAS analysis is also available 
\cite{Aad:2020sgw}
based on 140~fb$^{-1}$, addressing as well a fully hadronic final state.
They exclude a stop mass of approximately 1250 GeV for a massless 
$\tilde\chi^0_1$, as compared to the 1150~GeV for the CMS analysis,
but the exclusion tables provided cover a smaller range of stop
masses, up to 1450 GeV, and they have a coarser granularity. We therefore 
use the results of \cite{Sirunyan:2019ctn} for \ttbar + \etmiss as well,
although the limits are not the best available ones. 
Alternative CMS analyses  \cite{Sirunyan:2019glc, Sirunyan:2019xwh} 
exclude a stop mass of 1200~GeV,
but they do not provide results in tabular form at the time of writing.

For each $D_{\lambda,11}$ value considered,
the expected cross-section for each signature  as a function of $m_{\phi}$
is compared to the
mass-dependent excluded cross-section
from \cite{Sirunyan:2019ctn}, and the mass value
where the two curves cross is taken as excluded mass for that
configuration of couplings.
\begin{figure}[t]
\begin{center}
\includegraphics[width=0.45\textwidth]{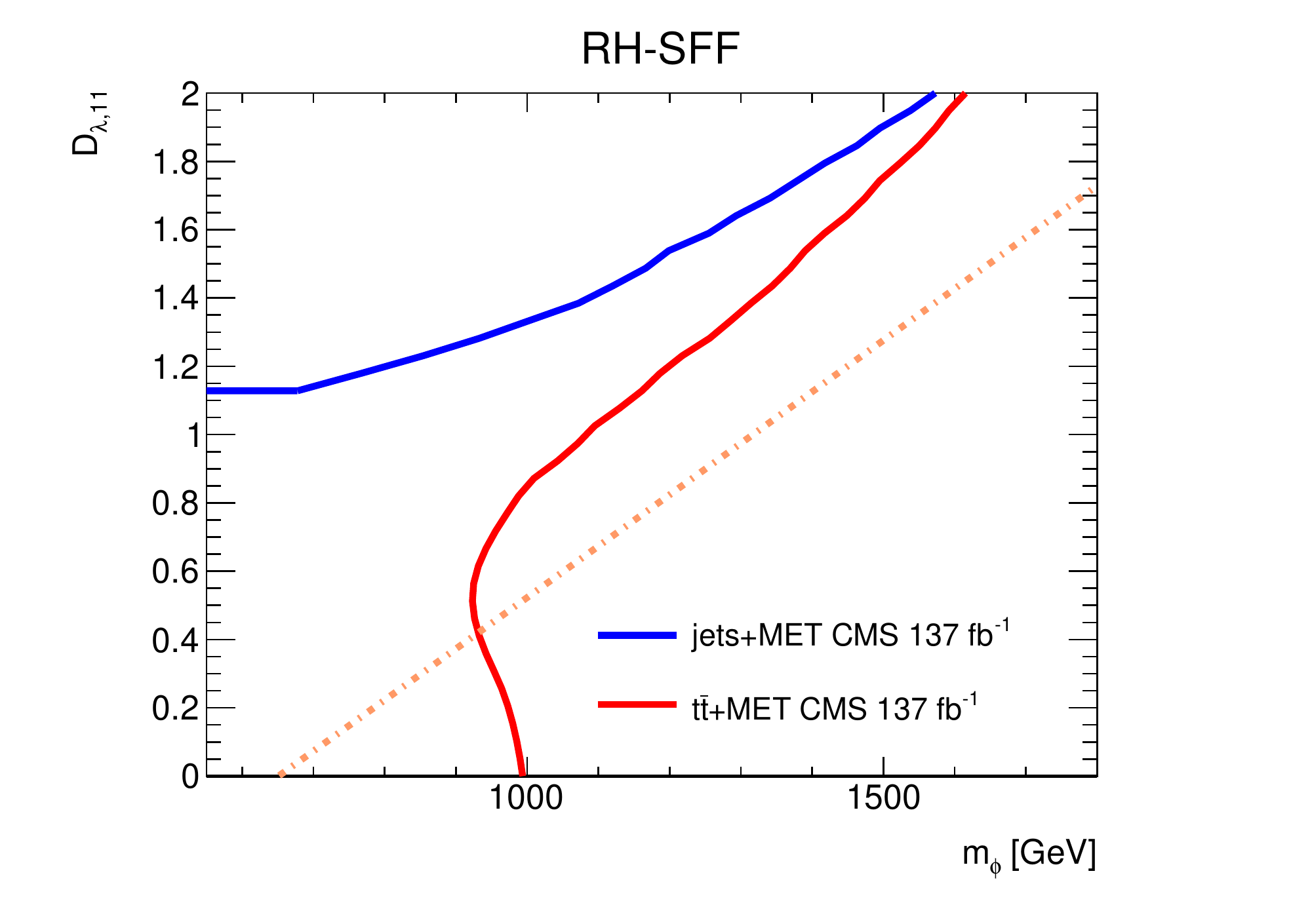}
\includegraphics[width=0.45\textwidth]{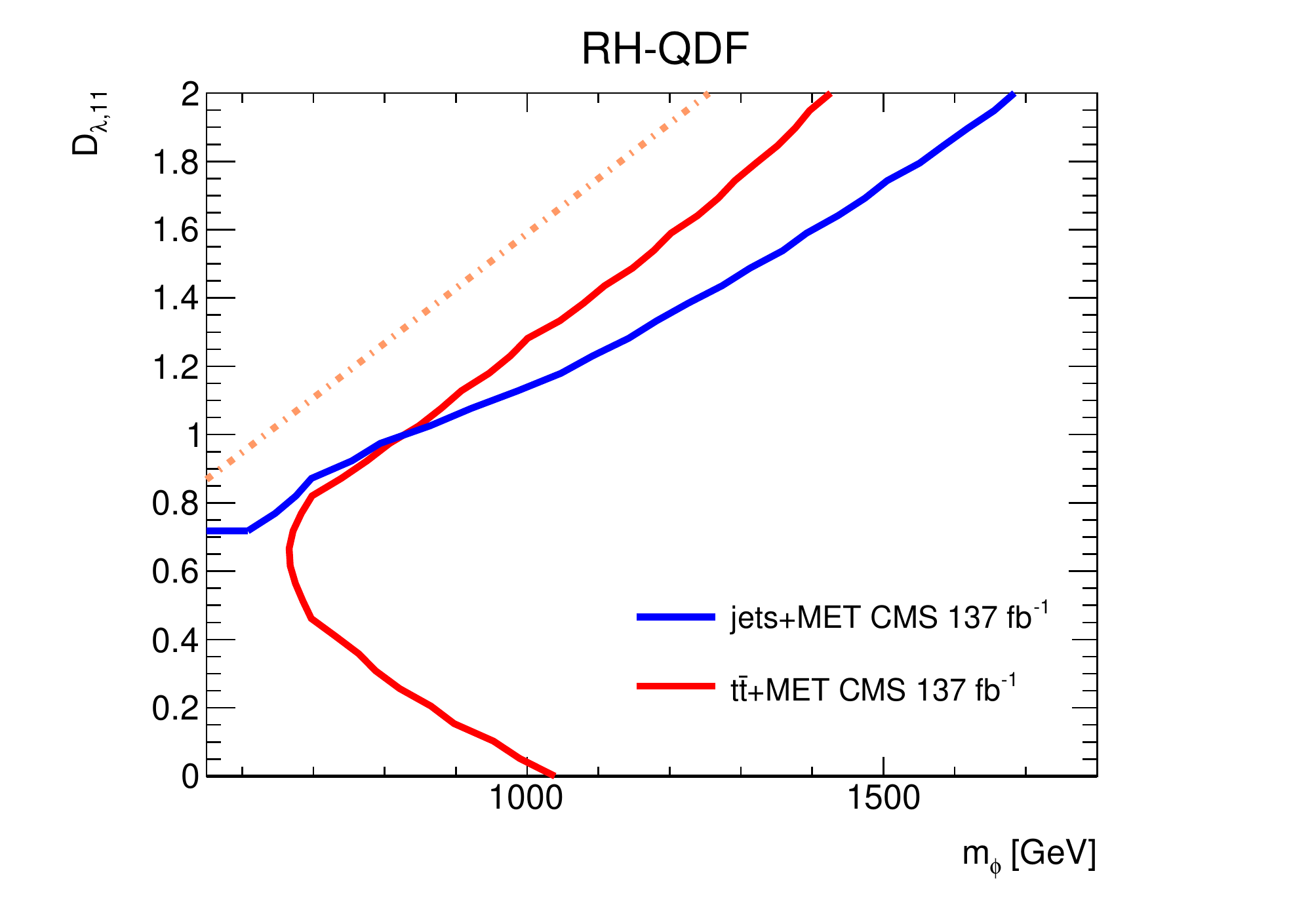}
\includegraphics[width=0.45\textwidth]{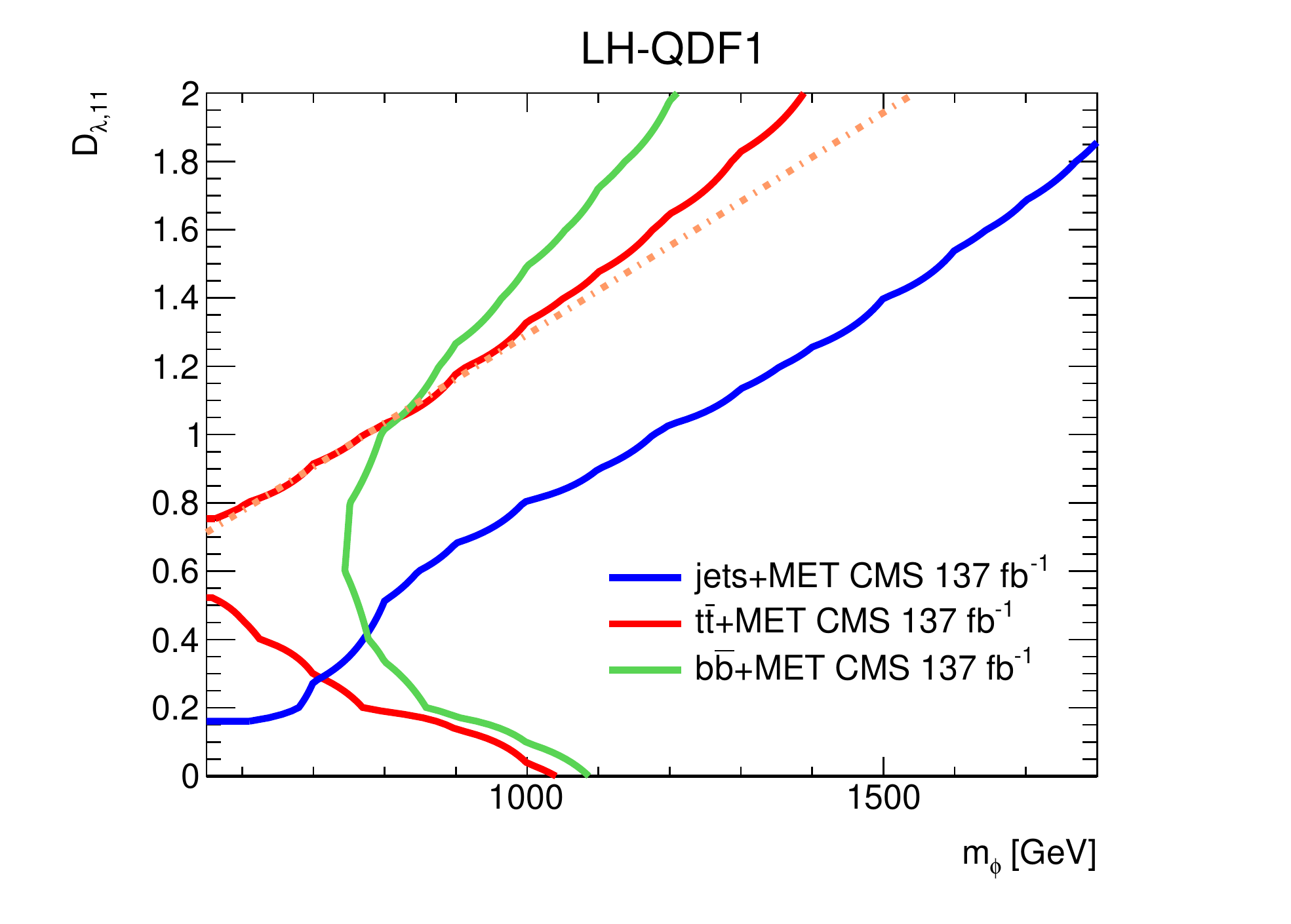}
\includegraphics[width=0.45\textwidth]{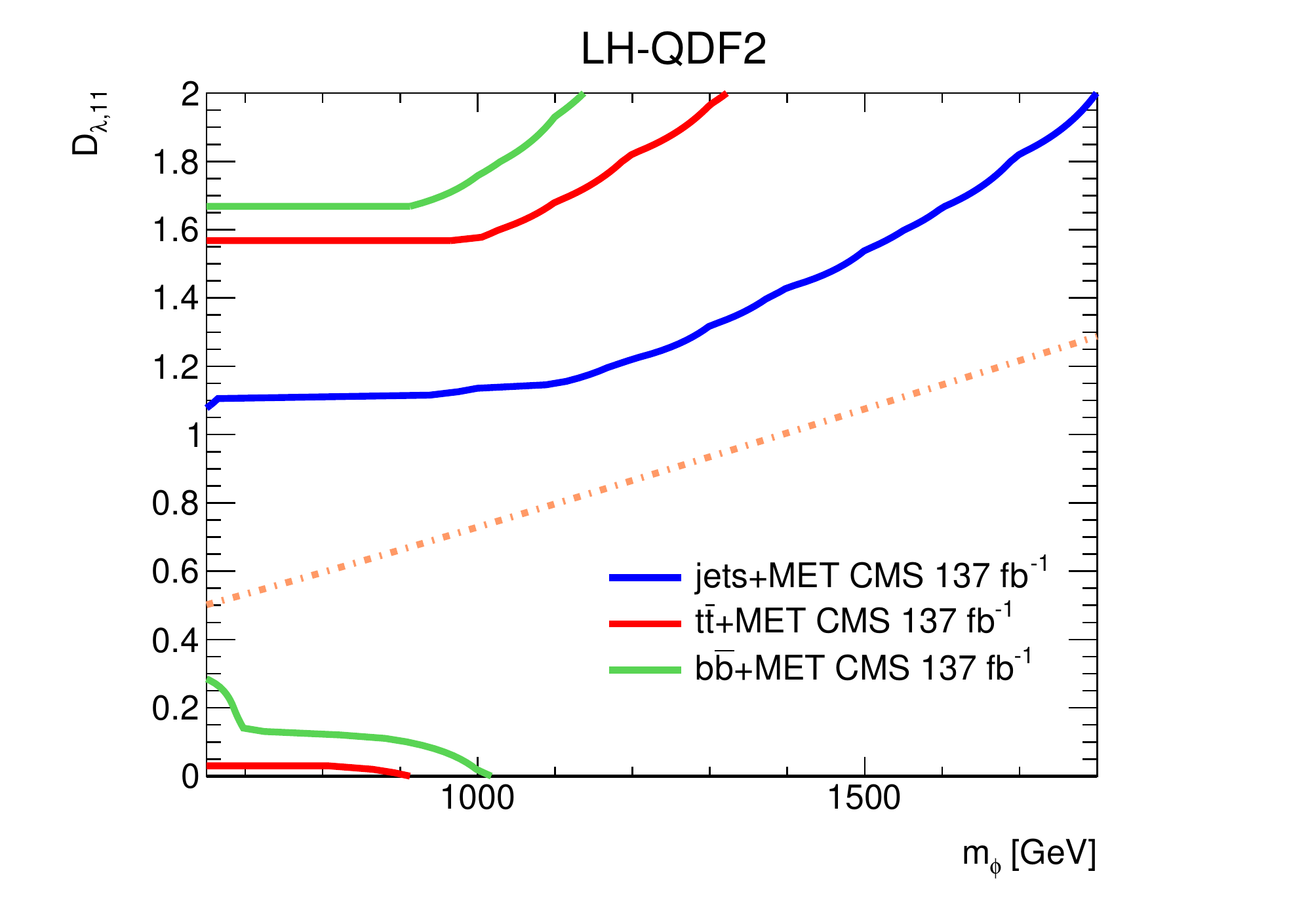}
\caption{Excluded areas in the ($m_\phi$, $D_{\lambda,11}$) plane based on the analyses of the CMS Collaboration on the full Run 2  statistics 13 TeV LHC data for the four benchmarks models proposed in this paper.
The excluded areas are on the left of the curves. {The orange dash-dotted lines indicate for which set of parameters the correct relic abundance is obtained.}}
\label{fig:recast}
\end{center}
\end{figure}
The results of the exercise are shown in Figure~\ref{fig:recast}. 
For the RH-SFF benchmark, where the $D_{\lambda,33}$ coupling 
is much larger than $D_{\lambda,11}$, the \ttbar+\etmiss signature 
is dominant over the whole considered {parameter} space, and $m_{\phi}$ lower than
900 GeV are excluded for all values of $D_{\lambda,11}$.
For the RH-QDF benchmark, where the first and third-generation
couplings have a comparable value, \ttbar+\etmiss dominates
at low $D_{\lambda,11}$ and $jj$+\etmiss at high $D_{\lambda,11}$,
and {values of $m_{\phi}$ up to 650~GeV are excluded.}
Concerning LH-QDF1, which has a similar pattern of couplings but the additional contribution of {down-type} quarks, $b\bar{b}$+\etmiss 
is dominant at low $D_{\lambda,11}$, with very similar power as
\ttbar+\etmiss, but $jj$+\etmiss again provides the best exclusion 
for high $D_{\lambda,11}$. The interplay of $b\bar{b}$+\etmiss and $jj$+\etmiss
brings the excluded value of $m_{\phi}$ to 800 GeV.
A similar pattern is observed for LH-QDF2,
but the very high value of  $m_{\chi}$ strongly reduces the
analysis acceptance for low mediator masses, leaving a large interval 
of $D_{\lambda,11}$ for which no value of {$m_{\phi}$} is excluded.
Dedicated analyses targeting models with small mass differences
between squark and neutralino are needed to improve the 
sensitivity in that area. 

{The orange dash-dotted lines in Figure~\ref{fig:recast} indicate for which set of parameters the correct relic abundance is obtained, assuming thermal freeze-out. To the left of the line, an additional DM component is required to explain the observed DM density. To the right of the line, thermal freeze-out leads to a too large DM density, an extension of the model would hence be required to avoid an overclosing universe. 
The minimal model with the correct relic abundance is excluded by $jj$+\etmiss for the LH-QDF benchmark, and by both $jj$+\etmiss and \ttbar+\etmiss for the RH-QDF benchmark. The simple thermal freeze-out assumption is however still viable in RH-SFF, due to the large splitting between the DM couplings to the different quark flavours, and in LH-QDF2, due to the large DM mass $m_\chi = 450$~GeV.} 

\section{Detailed analyses of single-top signatures}
\label{sec:ana}

For each of the three single-top signatures addressed {in section~\ref{sec:xsec}},  we perform a detailed study of the LHC prospects, focusing in each case on the semileptonic 
decay of the top quark. The final states of interest will then always
include an isolated lepton ($e$,$\,\mu$), a hadronic jet tagged as the
result of the fragmentation of a $b$-quark, and \etmiss both from 
the neutrino from the top decay and from the production of two $\chi$ particles
which escape the detector undetected. We will show in the following how, 
for each of the considered signatures, kinematic cuts can be defined 
to reduce to a manageable level the very large backgrounds 
from \ttbar and $W$+jets productions.

\subsection{Monte Carlo simulation}

All the samples are generated for a centre-of-mass energy of
the LHC of 14~TeV. 
The signal samples  are generated at LO using  the DMFV 
UFO model~\cite{Degrande:2011ua} implementations provided in
\cite{Blanke:2017tnb,Blanke:2017fum}. Parton-level events are 
generated with {\tt MadGraph5\_aMC@NLO}~\cite{Alwall:2014hca},
employing {\tt NNPDF3.0} parton distribution functions (PDFs)~\cite{Ball:2014uwa},
and showered with {\tt PYTHIA~8.2} \cite{Sjostrand:2014zea}.

For each of the three benchmarks, we generate a string of event samples 
with mediator mass variable between 400~GeV and 3~TeV and fixed values of the 
couplings  $D_{\lambda,11}=D_{\lambda,22}=0.4$. 
A grid of event samples with fixed mediator mass $m_{\phi}=1$~TeV and
$D_{\lambda,11}=D_{\lambda,22}$ variable between 0 and 2 
is generated in order to evaluate the dependence of the experimental 
acceptance on the different  sample compositions over the 
($D_{\lambda,11},D_{\lambda,22}$) plane.

The proposed analyses address final states with a single isolated 
lepton, therefore all of the Standard Model processes featuring a lepton 
in the final state are considered for the background evaluation. 

Backgrounds either with fake electrons from jet misidentification
or with real non-isolated leptons from the decays of 
heavy-flavour hadrons are not considered in this study.
An understanding of detector effects beyond
the scope of this study would be needed for a reliable estimate.
We estimate, based on the results of the ATLAS experiment, 
these backgrounds not to exceed around $15\%$ of the total
background surviving our selections.

Backgrounds from $\ttbar$~\cite{Campbell:2014kua},
$tW$~\cite{Re:2010bp}, $WW$, $WZ$ and $ZZ$ production~\cite{Melia:2011tj,Nason:2013ydw}
are generated at next-to-leading order (NLO) with {\tt POWHEG~BOX} 
\cite{Alioli:2010xd}. Samples of 
${\rm jets}+Z$ and ${\rm jets}+W$ events are generated at~LO with
{\tt  MadGraph5\_aMC@NLO}
and considering up to three jets for the matrix element calculation.
The $\ttbar V$ backgrounds with $V = W,Z$ are also simulated 
with {\tt  MadGraph5\_aMC@NLO} at LO with a multiplicity
of up to two jets, and the $tZ$ and $tWZ$ backgrounds at LO.
All partonic events are showered with {\tt PYTHIA~8.2}.
The samples produced with {\tt POWHEG~BOX} are normalised to the NLO cross
section given by the generator, except $t\bar{t}$ which is normalised to
the  cross section obtained at next-to-next-to-leading order (NNLO)
plus next-to-next-to-leading logarithmic
accuracy~\cite{Czakon:2011xx,Czakon:2013goa}. The ${\rm jets} + W/Z$ samples
are normalised to the known NNLO cross sections~\cite{Anastasiou:2003ds,Gavin:2012sy},
while the~$\ttbar V$~samples are normalised to the NLO cross-section as calculated
by  {\tt  MadGraph5\_aMC@NLO}.

The analysis is performed based on the following objects:
leptons ($e$, $\mu$), hadronic jets and \etmiss.
Stable leptons produced in the decays of real $W$ and $Z$ 
and isolated from hadronic jets are considered
in the analysis. Jets are built out of the momenta of all the stable particles 
depositing energy in the calorimeter except muons using an
anti-$k_t$ algorithm~\cite{Cacciari:2008gp}
with a parameter $R=0.4$, as implemented in  {\tt FastJet}~\cite{Cacciari:2011ma}. Jets originating from the hadronisation of bottom-quarks ($b$-jets) are experimentally tagged with high efficiency (\btagged\ jets).
The \ptmiss vector with magnitude \etmiss 
is built out of the transverse momenta of all the 
invisible particles in the event. 

The experimental effects are simulated by smearing the momenta of
the analysis objects and by applying efficiency factors where
applicable. The smearing and efficiency functions used to this purpose 
are tuned to reproduce the performance of the 
ATLAS detector ~\cite{Aad:2008zzm,Aad:2009wy}. 
The working point for jet $b$-tagging has an efficiency of 77\%, 
with a rejection factor  $\sim5$ and $\sim110$ respectively on $c$ 
and light jets. 

\subsection{Statistical procedures}
The LHC sensitivity to the three proposed signatures
is estimated for integrated luminosities of 140~fb$^{-1}$,
300~fb$^{-1}$ and 3000~fb$^{-1}$,
corresponding to the available statistics of LHC Run 2 and the 
projected statistics for LHC Run 3 and the high-luminosity LHC run 
respectively.
We assume the same detector performances for the high-luminosity LHC
as for the previous data-taking runs.

The sensitivity is calculated using a test statistics
based on a profiled likelihood ratio, and we make use of the CLs method~\cite{%
Read:2002hq} to obtain 95\% confidence level (CL) exclusion limits. The
statistical analysis is performed with the {\sc RooStat} toolkit \cite{%
Moneta:2010pm} and we assume systematic uncertainties of 15\% and 5\% on the SM
backgrounds and on the signal respectively. 

\subsection{Analysis strategy}

All of the addressed signatures, monotop, $tj$ and $tb$,  have a single top 
quark in the final state, and we consider its semileptonic decay. Therefore, the common basic selection for all
three signatures is the requirement of one and only one 
isolated lepton ($e$ or $\mu$) with $p_T>30$~GeV within $|\eta|<2.5$.
Considering the monotop signature, the signal has a strong imbalance 
in favour of a positive {lepton}, therefore only leptons 
with positive charge are considered. For all signatures, {we further require} the presence
of at least one $b$-tagged jet and of \etmiss from
neutrinos and dark matter escaping the detector. The requirements
on additional jet activity depend on the signature. For monotop and $tj$,
one and only one $b$-tagged jet is required, while for $tb$ two $b$-tagged jets
are required. Additional jets with $p_T>50$~GeV are vetoed for 
monotop and $tb$, whereas an additional high $p_T$ jet in the event is
expected for $tj$, from  the decay of $\phi$ into a light jet.

For events featuring a semileptonic decay of the top, the 
invariant mass of the lepton and the $b$-tagged jet \mbl 
has a sharp end-point around 160~GeV. For the $tb$ analysis, 
where two $b$-tagged jets are required, the minimum of the
two $b$-lepton invariant mass combinations has an equivalent property.
An upper limit at 160 GeV on this variable ensures a significant
reduction of non-top backgrounds.

The above requirements will select also {an overwhelming number of background events from} standard model processes
featuring {the} production of a $W$ boson decaying into leptons, 
dominated by \ttbar and $W$+jets production.
The main handle against these backgrounds is the fact that they have
\etmiss only from the neutrino from $W$ decay, whereas the signal 
has a large additional \etmiss from the invisible $\chi$ particles.
A useful variable to exploit this feature
is built out of the transverse momentum of the lepton
and \etmiss as:
\begin{equation}
  \mTlep \!\equiv\!\sqrt{2\,|\ptl|\,\etmiss\,(1\!-\!\cos\Delta\phi(\ptl,\ptmiss))},
\end{equation}
where \ptl is the transverse component of the momentum of the lepton, and \ptmiss is defined in the previous section. For events where the lepton and all of the \etmiss are produced in the decay of a single $W$, this variable is bounded
from above by the $W$ mass.  For the $tj$ and $tb$ signatures a lower 
limit of 160 GeV on this variable reduces then the single $W$ backgrounds
to a manageable level. For monotop, which is a simpler final state
with less kinematic handles to reduce the backgrounds, a stronger limit
of {250~GeV} on \mTlep is applied.

The dominant background after an appropriate $\mTlep$ requirement
is typically composed of  \ttbar events where both tops decay semileptonically
and only one lepton is detected. The by now standard approach to
reject this background is the asymmetric
\mttwo variable (denoted \amttwo)~\cite{Konar:2009qr, Lester:2014yga} that
consists in a variant of the \mttwo observable. The \amttwo variable is built
from two legs (corresponding to the two decay chains) containing both a visible
part and an invisible part, and it requires two test masses corresponding to the
invisible mass attached with each leg. For the \ttbar 2-lepton background with
one lepton lost,  the visible part of the first leg is
the vector sum of the momenta of the $b$-tagged jet and of the lepton, with
a test mass  set to zero. The visible part of the
second leg is built choosing among  the additional jets in the
event the one  with the highest $b$-tagging weight,
and {the} test mass is set to 80 GeV. Of course a selection on this 
variable cannot be applied for monotop, where only the $b$-jet
from the decay of the top is allowed in the event.

After removing the background events within the $W$ kinematic bounds,
there is still a significant background from events where a large \etmiss
is produced by the mismeasurement of a jet. These events have the
\etmiss aligned with the momentum of a jet. The $\Delta\phi_{min}$ variable is built
as the minimal angular difference in the transverse plane between \etmiss
and any reconstructed jet in the event. This variable has an enhancement 
near zero for the background, and a lower limit on it increases the signal over background ratio.

The angular differences in the plane transverse to the beam 
between pairs of observed objects can provide discrimination 
power between signal and background. The variables used in the analysis are
$\Delta\phi_{b\ell}$ and $\Delta\phi_{m\ell}$, the angular difference of 
the lepton  {with the $b$-jet and \etmiss, respectively.}
For the $tb$ analysis the former variable is {built} with the $b$-jet
most likely to be from the top decay. Harder cuts on the angular variables
are applied for the monotop signature with respect to $tj$ and $tb$,
again to compensate for the less constrained kinematics of the signature.

Finally, the $tj$ analysis addresses the signal topology 
featuring one leg where a heavy mediator decays into a hard jet 
and dark matter. In this case
the distribution in the transverse mass
built from the transverse momentum of the hard jet  and the one of the
$\chi$ particles has an end-point at  $(m_{\phi}^2-m_{\chi}^2)^{1/2}$.
This feature can be exploited by defining a dedicated variation on \amttwo
which we call the \mttwoblj variable.
The visible part of the first leg is built from the sum of the momenta of
the $b$-tagged jet and of the lepton, and the test mass is set to zero.
The visible part of the second leg uses the hardest non-$b$-tagged
jet, and again a zero test mass.

The selection criteria for the signal regions for the three analyses,
based on the variables defined above, are given in Table~\ref{tab:cuts}.
\begin{table}[h!]
\begin{center}
\small
\begin{tabular}{p{2.5cm}ccc}
\toprule
	Variable [Unit]      &   monotop  &  $tj$ & $tb$    \\
\midrule
$N_{\ell}$   &  $=1$  &   $=1$   & $=1$     \\
$p_T(\ell)$  [\GeV]    & $>30$   &  $>30$   & $>30$      \\
charge($\ell$)    & $>0$   &  any   & any      \\
$N_{b-jet}$                & $=1$   &   $=1$   & $=2$      \\
$p_T(b_1)$  [\GeV]    & $>30$   &  $>30$   & $>110$      \\
$p_T(b_2)$  [\GeV]    &  -  &   -   & $>110$    \\
$p_T(j)$  [\GeV]      & $<50$   &   $>100$   &  $<50$     \\
\mbl [\GeV]            &  $<160$  &   $<160$   & $<160$ \\ 
\mTlep [\GeV]            &  $>250$  &   $>160$   & $>160$ \\ 
$\Delta\phi_{min}$  [rad]  & $>2$   &  $>0.6$  &  $>0.7$ \\
$\Delta\phi_{b\ell}$ [rad]    & $<1.5$  &  incl &  $<1.5$ \\
$\Delta\phi_{m\ell}$  [rad]   & $>2.5$  &  incl. &  incl. \\
\etmiss  [\GeV]        &  $>400-600$      &   $>90$    &  $>250$   \\
\amttwo  [\GeV]        &   -       &   $>250$  &  $>300-500$   \\
\mttwoblj  [\GeV]        & -     &  $>300-500$   &  - \\ 
\bottomrule
\end{tabular}
\large
\caption{Summary of the selection criteria for the three proposed 
single-top analyses. The cuts on \etmiss and  \mttwoblj respectively
for the monotop and $tj$ analyses are optimised separately for 
each value of $m_{\phi}$ within the range given in the table.
The variable $p_T(j)$ is the transverse momentum of the hardest jet not tagged 
as a $b$-jet.}
\label{tab:cuts}
\end{center}
\end{table}
For all three signatures, the final sensitivity is calculated 
for each signal point considered by 
varying the cut value on the final discriminant variable, which 
is \etmiss for monotop, \mttwoblj for $tj$ and \amttwo for $tb$.

\begin{figure}
\begin{center}
\includegraphics[width=0.47\textwidth]{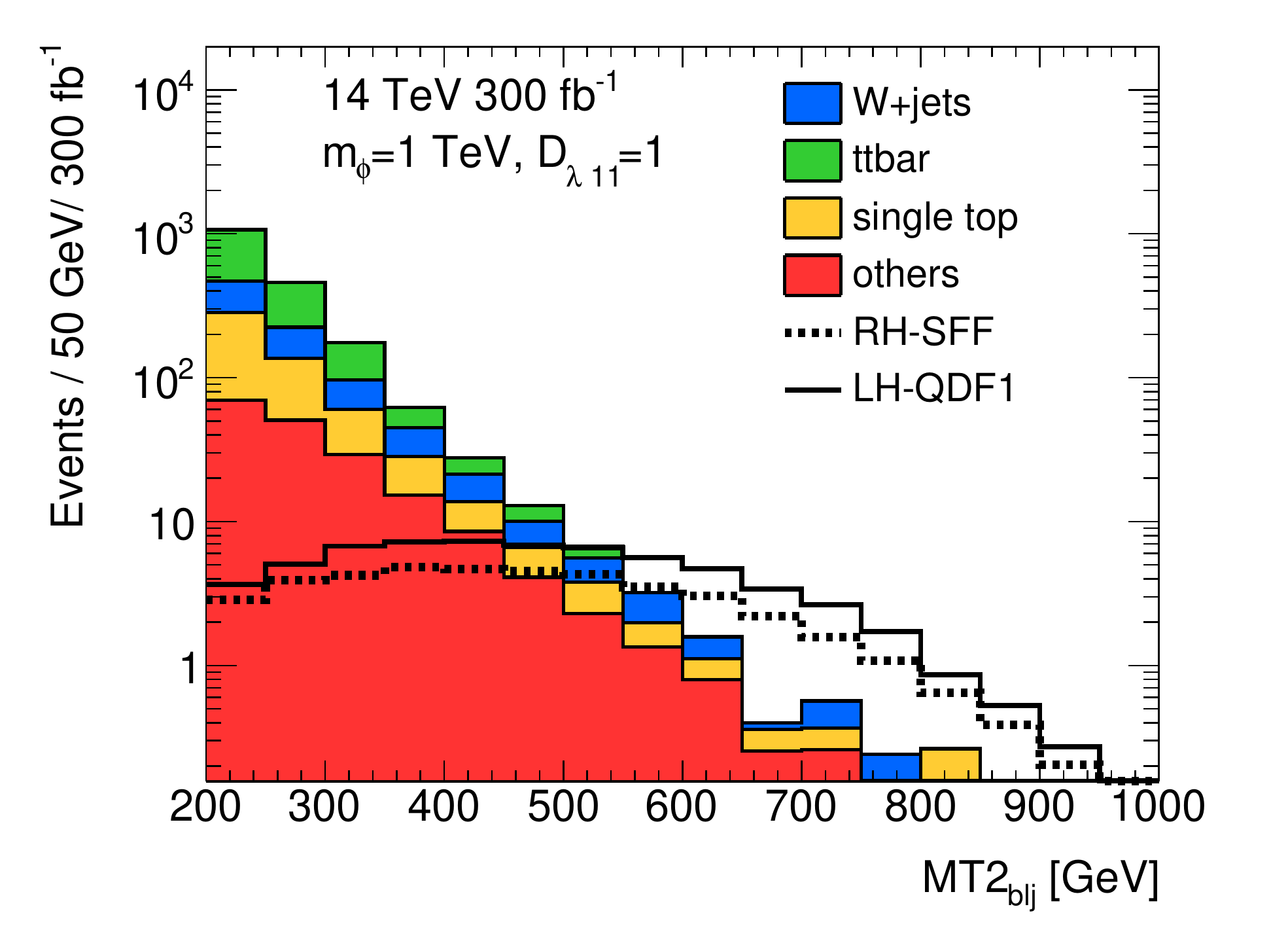}
\includegraphics[width=0.47\textwidth]{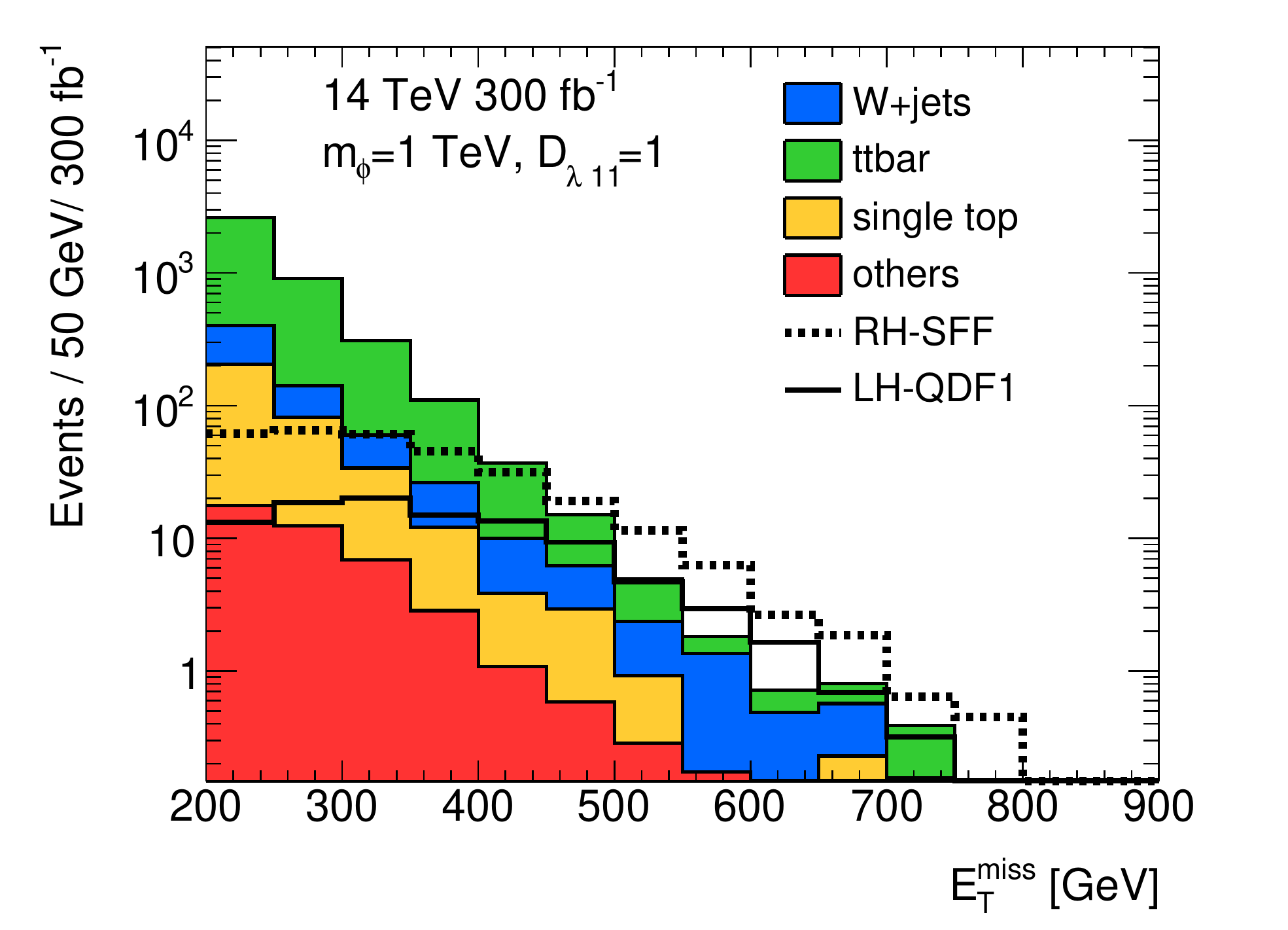}
	\caption{Distribution of the \mttwo (left) and \etmiss (right) variables for the $tj$ and monotop analyses respectively
for two signal points and for the SM backgrounds. The distributions 
are shown after all of the selection
cuts are applied except the one on the plotted variable. The normalisation
corresponds to an integrated luminosity of 300~fb$^{-1}$.}
\label{fig:distri}
\end{center}
\end{figure}

The distributions of the final discriminant variable after all other
cuts have been applied are shown in Figure~\ref{fig:distri} for 
the backgrounds and for benchmark signal samples for the $tj$ and
monotop analyses. 

For $m_\phi=1$~TeV and a cut of 400~GeV on the final discriminant,
the signal efficiency is between 0.7 and 1\% for monotop, 
with a background of $\sim60$ events for 300~fb$^{-1}$.  
For $tj$, the efficiency is between 3 and 5\% and the background of $\sim50$ events. For $tb$, the efficiency is between 1\% and 1.5\% for a background of $\sim6$ events. 

{The efficiency of the analysis strategies outlined above is only mildly 
dependent on $m_\phi$, but it displays a characteristic threshold
dependence on the  mass difference  $\Delta m\equiv m_\phi-m_\chi$. 
For fixed $m_\phi$ and 
$\Delta m>700-800 \; \mathrm{GeV}$, the efficiency is approximately independent 
from $\Delta m$, whereas it quickly decreases when $\Delta m$ decreases
below the threshold value. For low $\Delta m$ the visible decay products
of $\phi$ become soft, failing the kinematic requirements which
are necessary to suppress to an adequate level the SM backgrounds. 
Therefore the results shown below will apply to different choices
of $m_\chi$ than the ones corresponding to our benchmarks 
for values of $m_\phi$  such that $\Delta m$ is above threshold.}

Based on these results and on the dependence of the production cross-section of the signatures on the model parameters, we can explore in the next section the relative importance of the different signatures over the parameter space of 
the model for each of the four benchmarks.

\section{Results}
For each of the proposed benchmarks we compare the
area excluded by the CMS analysis in the ($m_\phi$, $D_{\lambda,11}$)
plane based on 137~fb$^{-1}$ of 13 TeV LHC data to the 
area covered on the same plane by each of the single top analyses
proposed in this paper for 137 and 300~fb$^{-1}$ of 14 TeV LHC data.
The purpose is to verify whether the new analyses would 
provide a gain in coverage of the parameter space of 
the model with respect to the existing flavour-conserving
SUSY searches, and to gauge the expected improvement with 
the doubled luminosity expected at the end of the LHC Run 3. 
{We note that the sensitivities for 137~fb$^{-1}$ are shown 
for a 14 TeV LHC, whereas the CMS results are for 13~TeV, therefore
the comparison is not fully consistent. It is anyway useful 
to give an idea, with a comparable amount of data, of  the
relative power of the different analyses in different regions
of the parameter space.}

\begin{figure}[t]
\begin{center}
\includegraphics[width=0.47\textwidth]{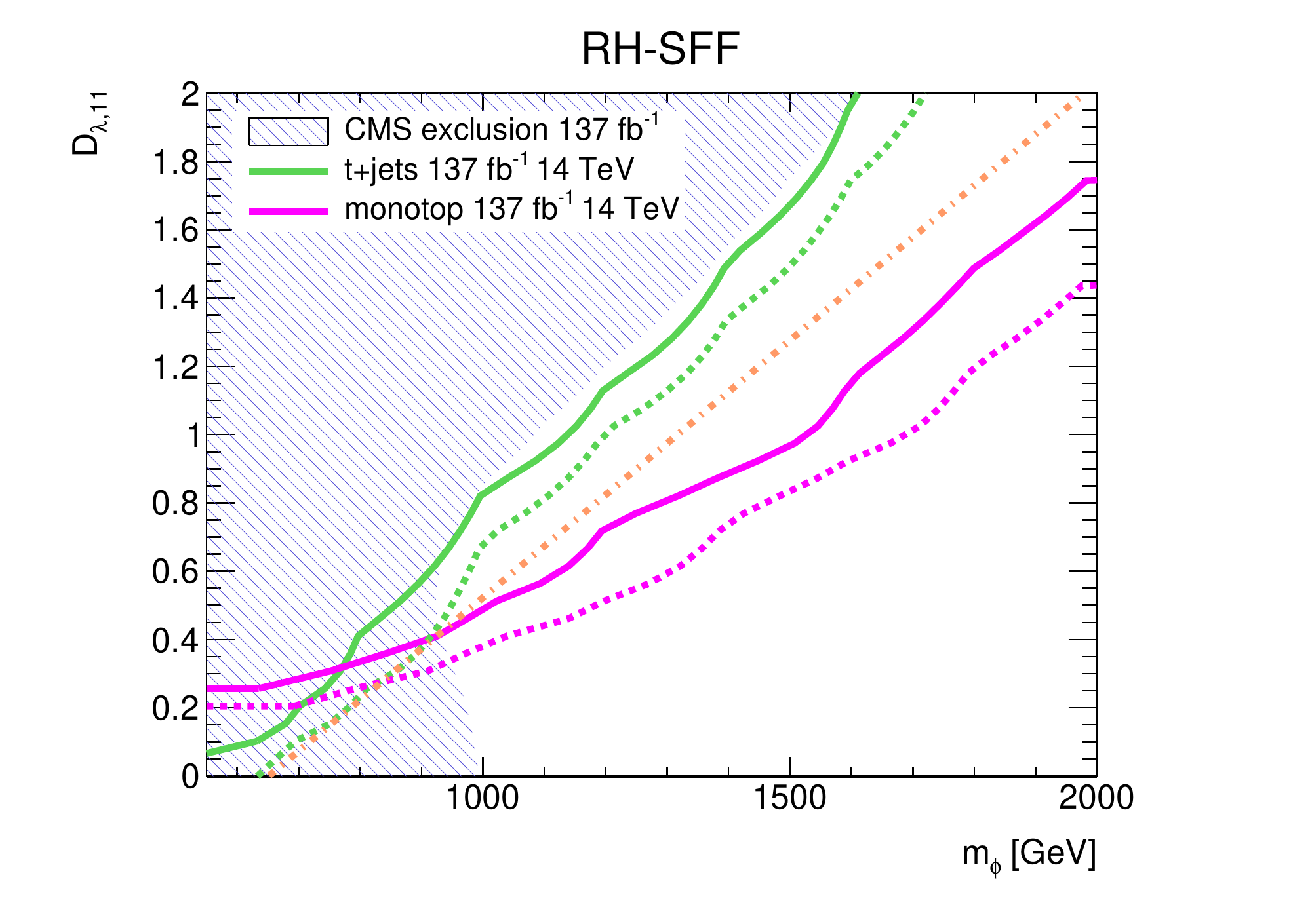}
\includegraphics[width=0.47\textwidth]{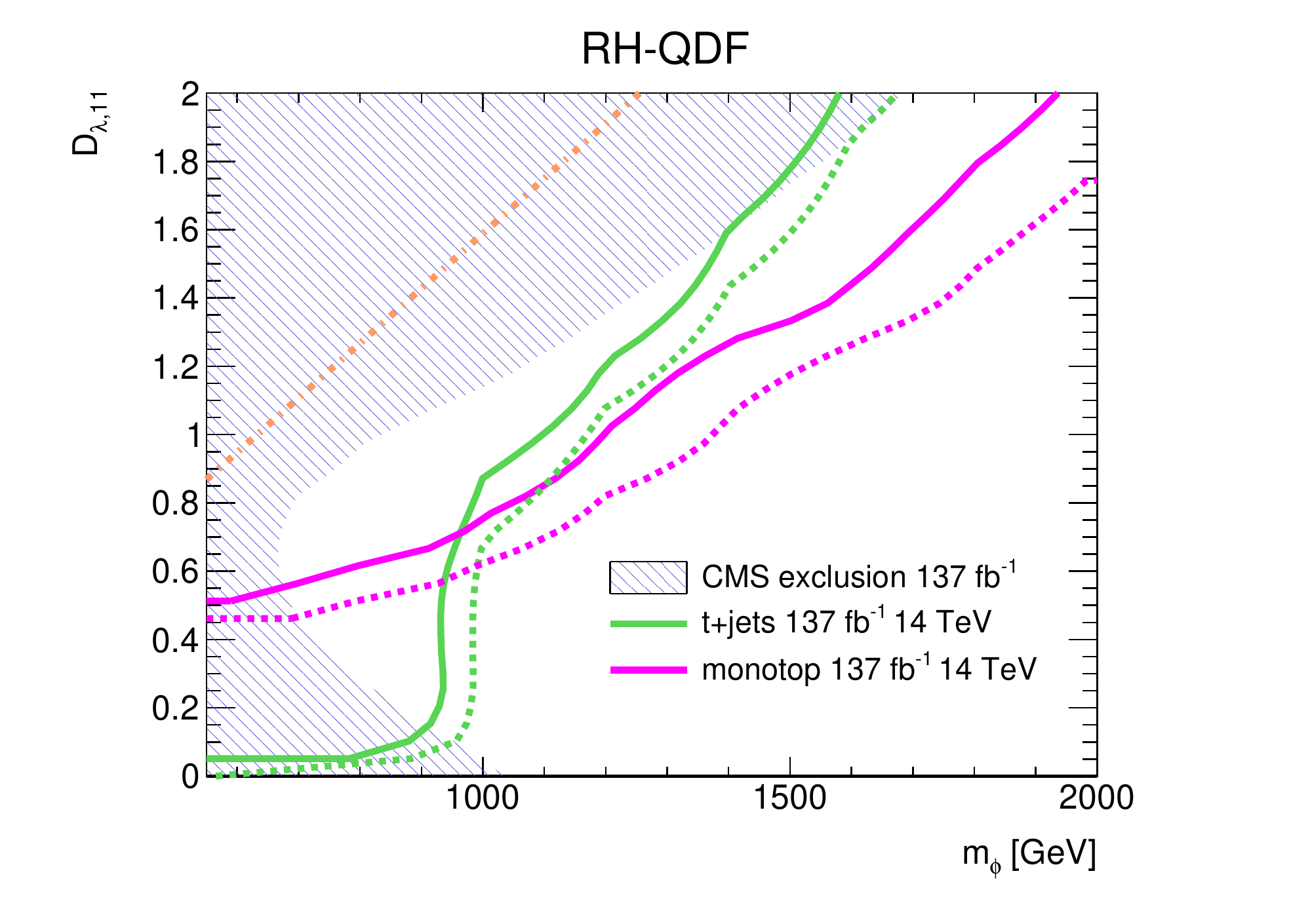}
\includegraphics[width=0.47\textwidth]{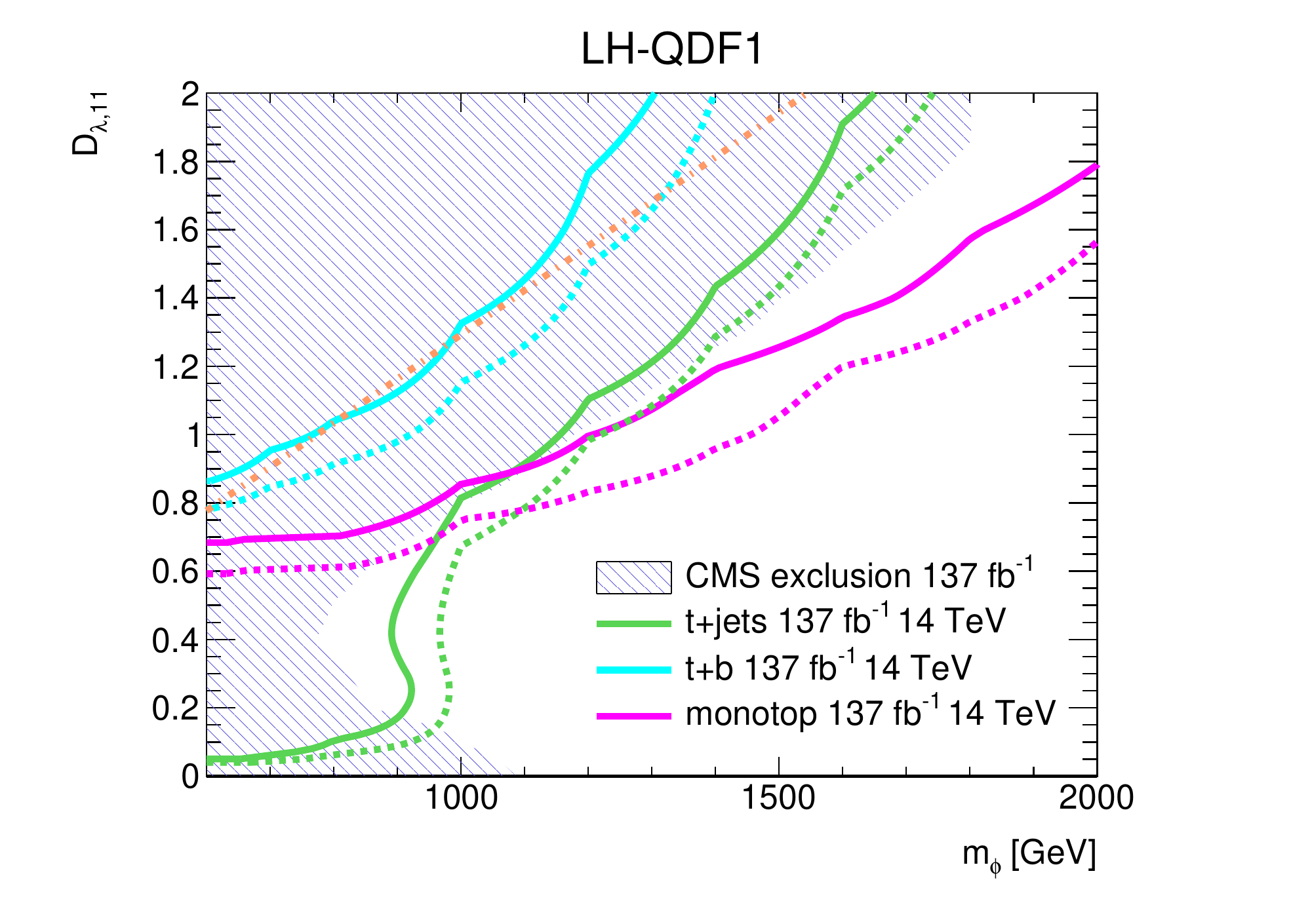}
\includegraphics[width=0.47\textwidth]{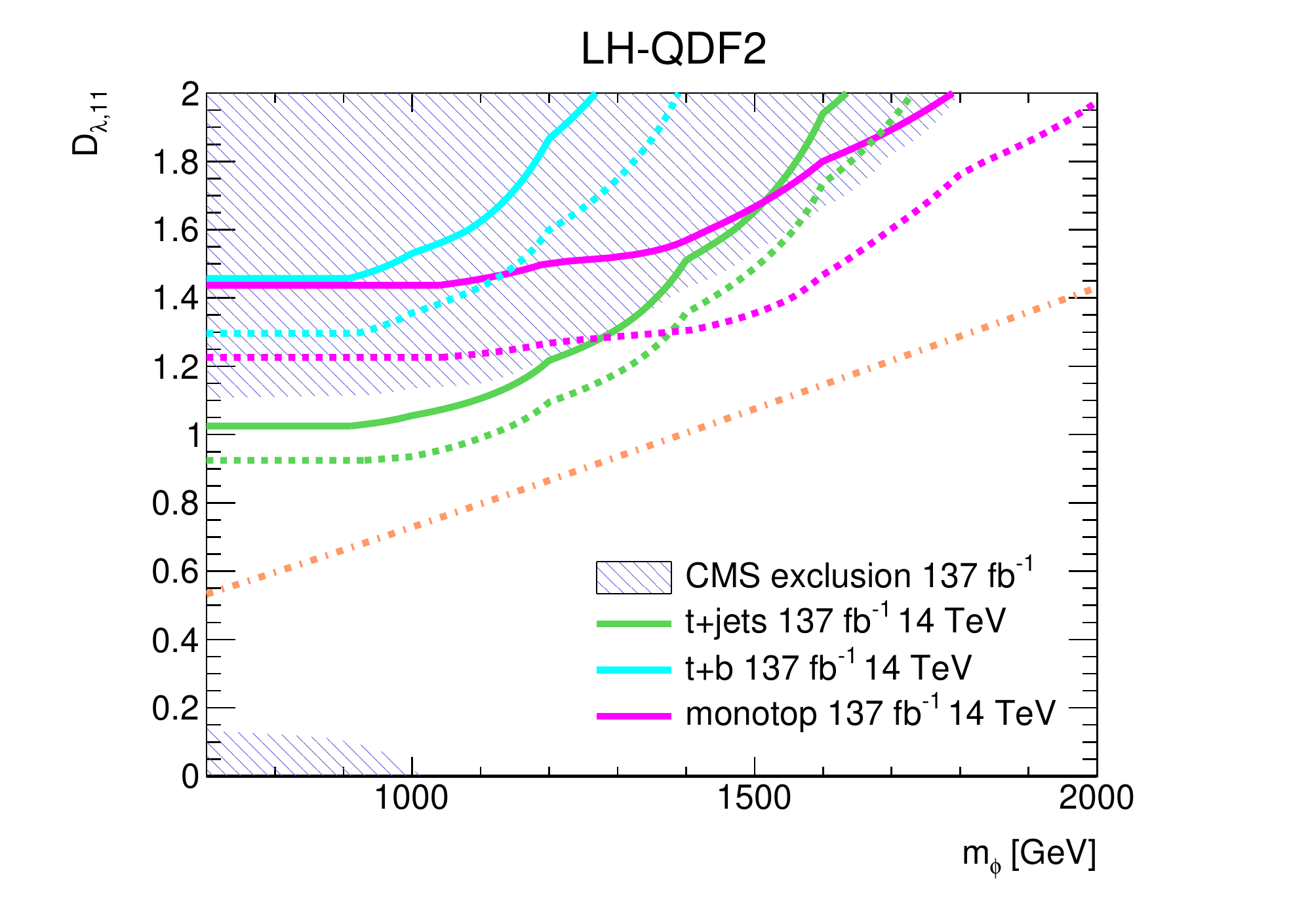}
	\caption{Full lines: excluded areas in the ($m_\phi$, $D_{\lambda,11}$) plane for the $tj$, $tb$ and  monotop analyses proposed in this paper for an integrated luminosity of  137~fb$^{-1}$ at a 14 TeV LHC.  The excluded areas are on the left of the curves.
The dashed lines show the corresponding results for 300~fb$^{-1}$. The
shaded area corresponds to the region excluded by the CMS analyses.  {The orange dash-dotted lines indicate for which set of parameters the correct relic abundance is obtained.}}
\label{fig:lim140all}
\end{center}
\end{figure}

The results are shown for the four benchmarks in Figure~\ref{fig:lim140all}, 
where the presently excluded region is shown as a shaded area, and 
the coverage of the single top analyses is shown as full (dashed) lines
for 137 (300) fb$^{-1}$.

In RH-SFF, the coupling to the third generation dominates over
the one to the first generation, and a dedicated monotop analysis would 
increase the reach in mediator mass  for $D_{\lambda,11}>0.4$, 
with the $tj$ analysis covering region similar to the one already
excluded by CMS for $D_{\lambda,11}>0.6$. {Already with 137 fb$^{-1}$ the monotop analysis would probe the parameter space predicted by thermal freeze-out.} For low values of $D_{\lambda,11}$,
none of the single top analyses would improve on the CMS limits from 
the stop analysis. 

For RH-QDF, where $D_{\lambda,11}$ and $D_{\lambda,33}$ have similar values,
the flavour conserving analyses display a minimum in the mass coverage 
for $D_{\lambda,11}\sim0.6$, corresponding to the situation where
the $tt$ and $jj$ signatures have both a 25\% BR.
The $tj$ signature for low $D_{\lambda,11}$ is dominated
by the doubly resonant production of the mediator, with one mediator 
decaying into a light quark and the other mediator decaying into top, 
with a 50\% BR approximately constant for $D_{\lambda,11}>0.2$,
as discussed in Section \ref{sec:pair-prod}.
As a result, the $tj$ analysis increases significantly the coverage
with respect to the flavour-conserving searches, with an approximately
flat mass reach of $\sim900$ GeV in the $D_{\lambda,11}$ interval 
from $\sim0.1$ to $\sim0.6$. For higher values of $D_{\lambda,11}$, the
$t$-channel production becomes dominant and the mass reach approaches the one of 
the jet-jet CMS analysis. The monotop analysis gives the best reach 
for $D_{\lambda,11}>0.7$, improving by several hundreds of GeV the
mediator mass reach with respect to the flavour-conserving analyses.

The benchmark LH-QDF1 is very similar to RH-QDF, with the main 
difference that the mediator in this case couples both with 
up and down quark flavours, thus altering the relative branching
ratios and opening up channels with $b$-jets in the final state.
The pattern of mass reach of  the different channels indeed 
approximately matches the one for RH-QDF1, with the $tj$ channel
increasing the mass reach in the region where the flavour-conserving
exclusion has a minimum, and with the dominance of monotop at 
high $D_{\lambda,11}$.
The additional $tb$ channel has sensitivity only in regions already
excluded by the CMS searches.

The phenomenology of the LH-QDF2 benchmark, as discussed above, 
is dominated by the high value of $m_{\chi}=450$~GeV.
The single top channels cover marginally the range of 
$D_{\lambda,11}$ couplings for which the existing analyses
have no sensitivity. The $tj$ analysis excludes 
the uncovered region with $D_{\lambda,11}$ between $\sim$1.0 and $\sim$1.1 
for the lowest allowed mediator masses. The monotop and the 
$tb$ analyses only cover the area already excluded by CMS.

The curves for 300~fb$^{-1}$ follow closely the ones 
for 137~fb$^{-1}$, with a gain between 0.1 and 0.3 units of 
$D_{\lambda,11}$ depending on the signature and on the mediator mass.
The monotop reach in LH-QDF2 is statistics limited, and the 
doubling of the statistics yields a somewhat larger improvement
than for the other channels/benchmarks.

\begin{figure}[h]
\begin{center}
\includegraphics[width=0.47\textwidth]{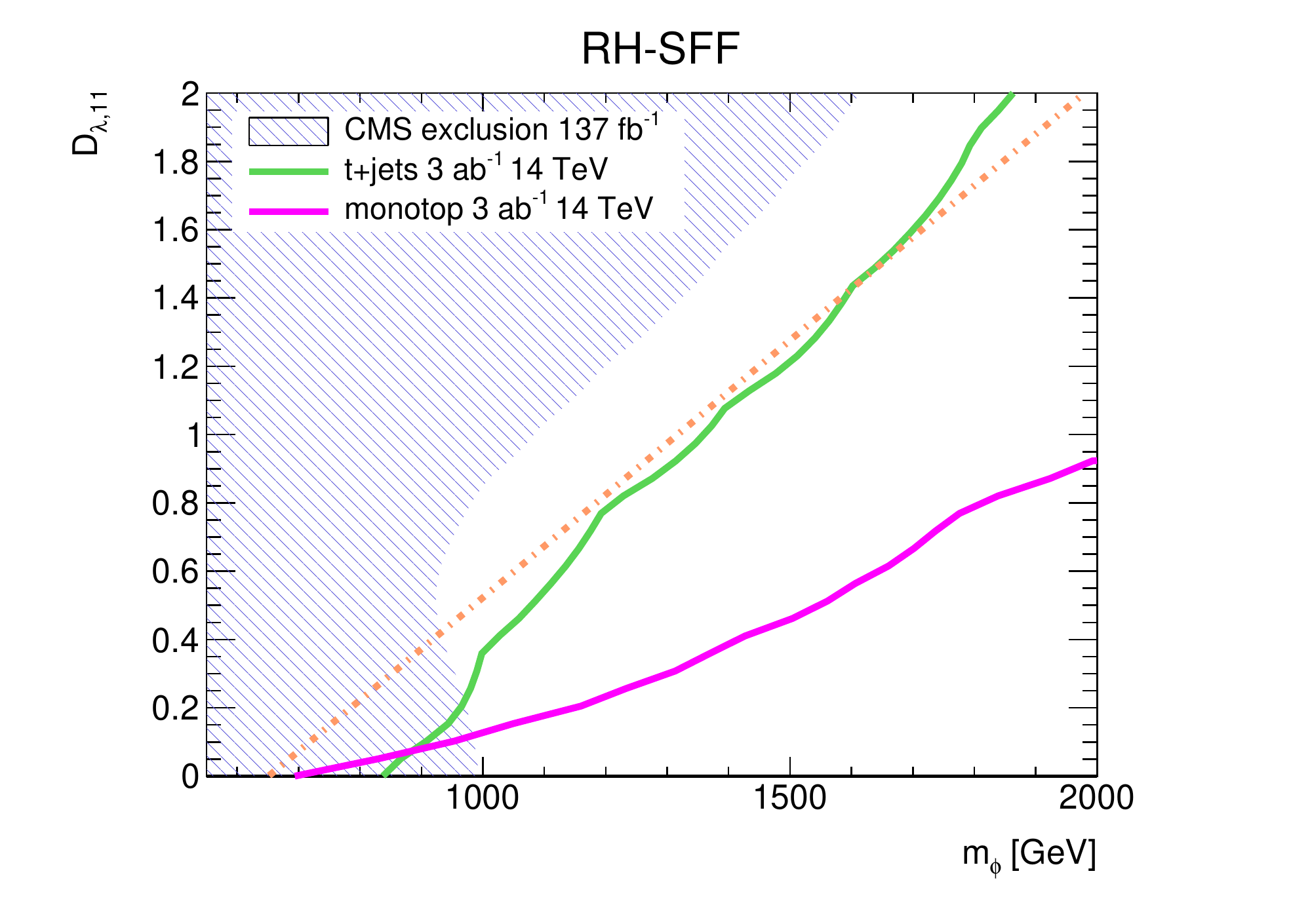}
\includegraphics[width=0.47\textwidth]{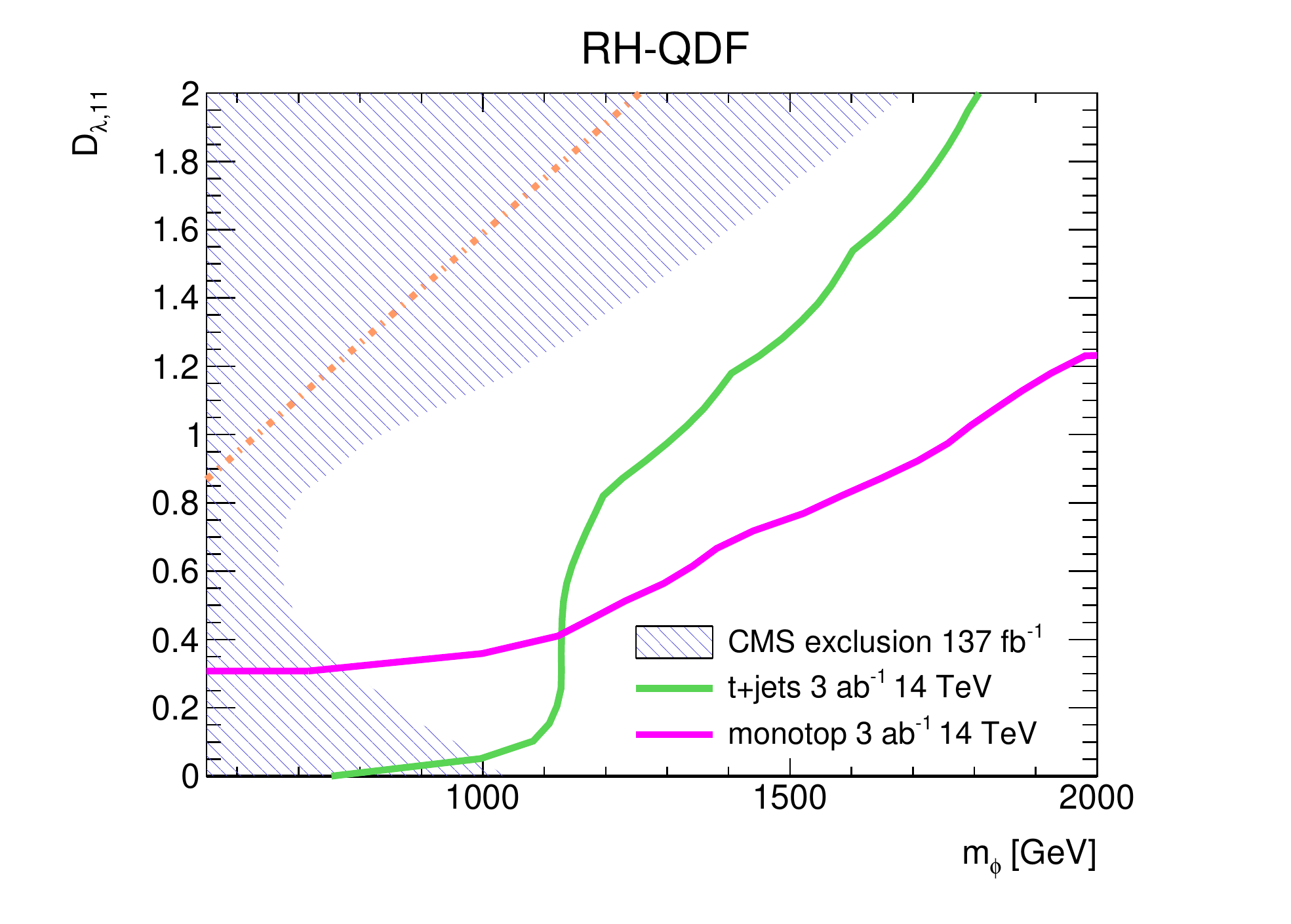}
\includegraphics[width=0.47\textwidth]{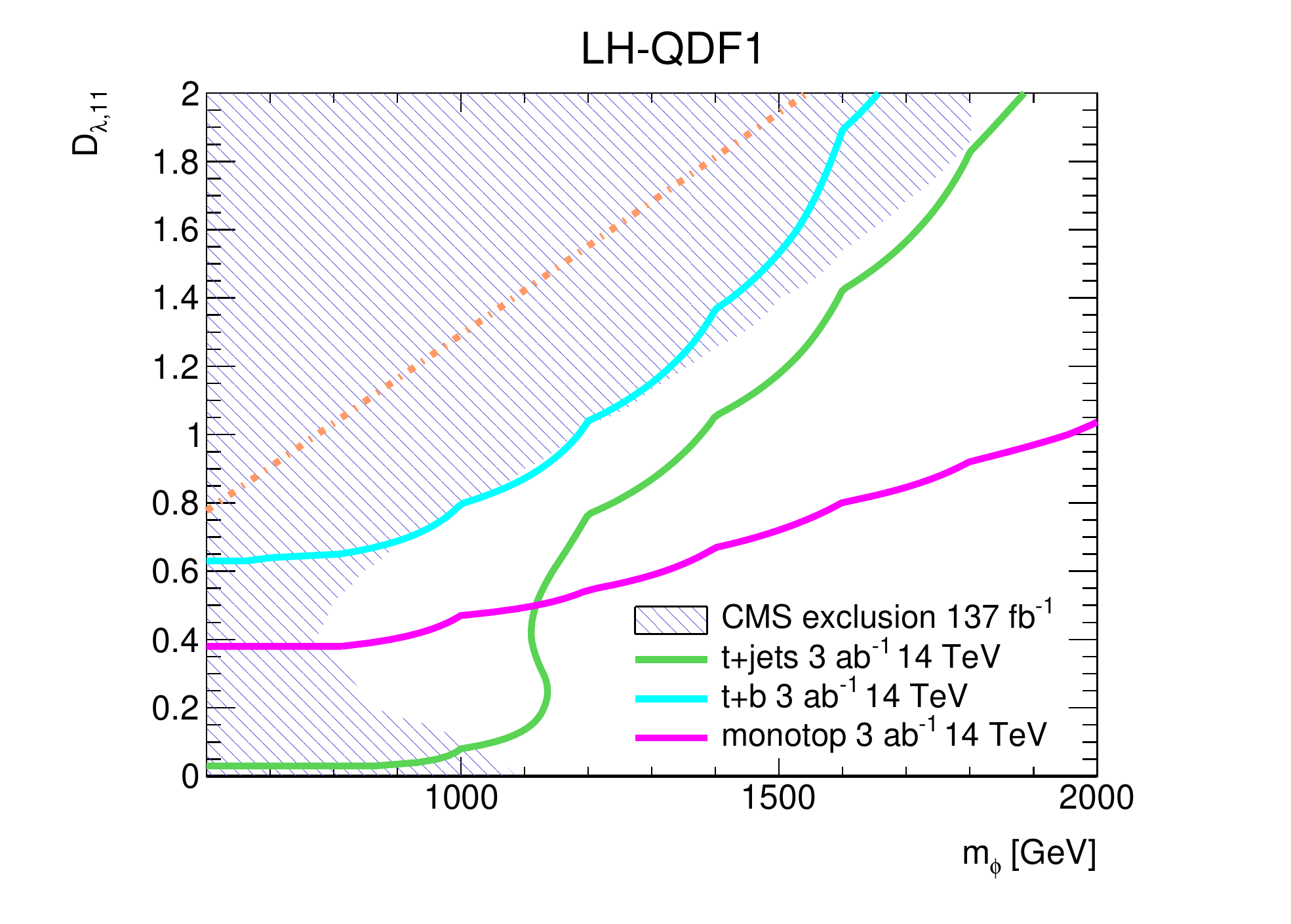}
\includegraphics[width=0.47\textwidth]{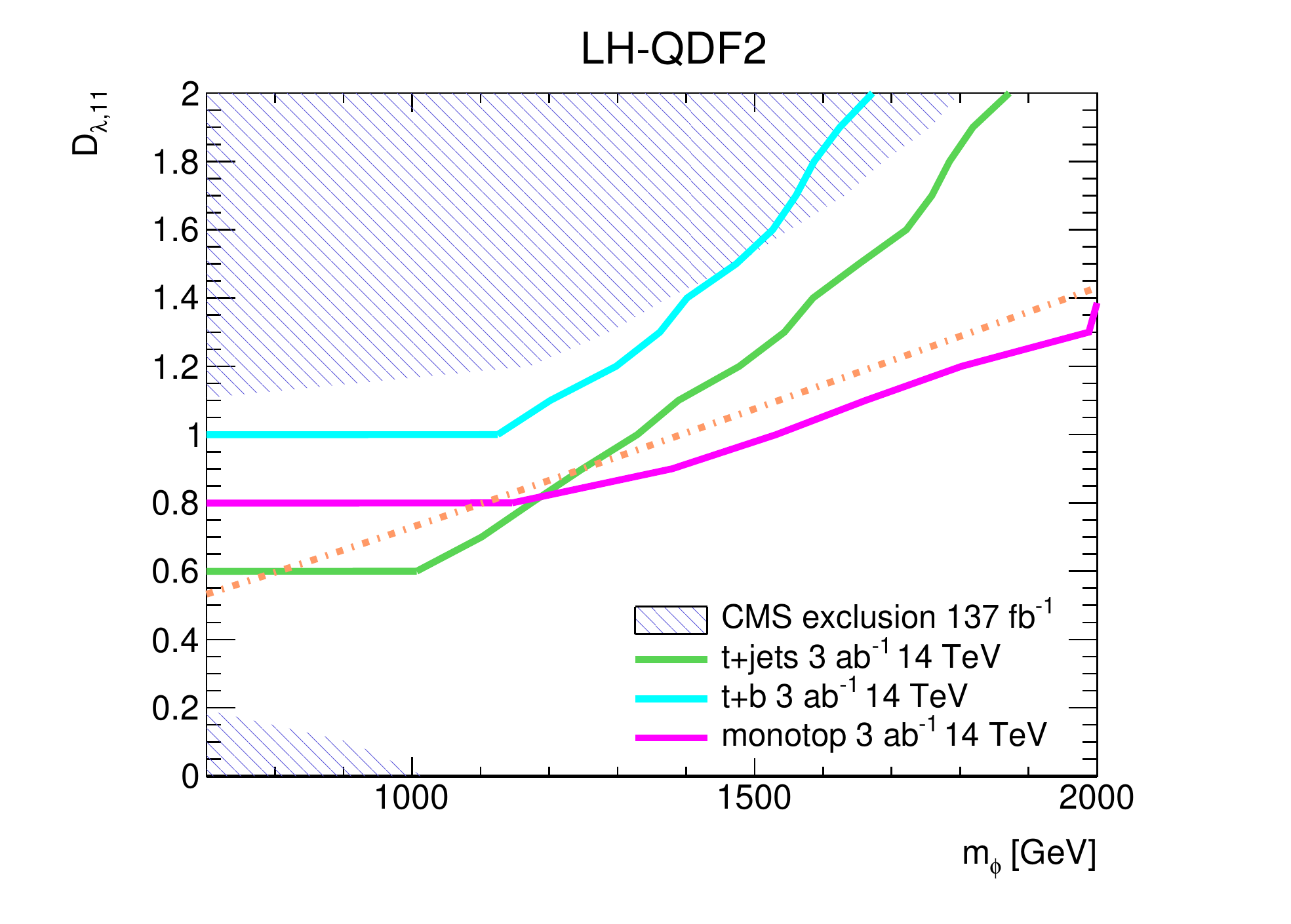}
	\caption{Excluded areas in the ($m_\phi$, $D_{\lambda,11}$) plane for the $tj$, $tb$ and  monotop analyses proposed in this paper for an integrated luminosity of  3~ab$^{-1}$ at a 14 TeV LHC.  The excluded areas are on the left of the curves. The
shaded area corresponds to the region excluded by the CMS analyses.  {The orange dash-dotted lines indicate for which set of parameters the correct relic abundance is obtained.}}
\label{fig:lim3000all}
\end{center}
\end{figure}
In Figure~\ref{fig:lim3000all} the statistics of the HL-LHC (3000 fb$^{-1}$) 
are shown,  based on the statistical procedure defined above.
For all of the benchmarks the monotop analysis will cover 
mediator masses of 2~TeV for $D_{\lambda,11}$ varying between 
0.8 and 1.2 dependent on the benchmark.
For lower values of the couplings, the impact of the
single top analyses depends on the specific benchmark.

Another common feature is that for values of 
$D_{\lambda,11}\leq0.1$ even at the HL-LHC the single
top analyses will not be able to improve on the CMS 
exclusion limit. In the low $D_{\lambda,11}$ region the 
dominant sensitivity will always be provided by the SUSY stop
searches. 

For RH-SFF, the monotop search will always provide the best
mass coverage among the proposed analyses, but it will
compete with the HL-LHC stop searches for low $D_{\lambda,11}$.

In the case of RH-QDF and LH-QDF1, the excluded mediator mass in the 
region between $D_{\lambda,11}=0.1$  and $D_{\lambda,11}=0.8$, which
are difficult for {flavour-conserving} analyses, 
will be $\sim1150$~GeV,  but monotop will provide better coverage
starting from $D_{\lambda,11}\geq0.5$.

The uncovered region in $D_{\lambda,11}$ for LH-QDF2
will be reduced to the interval between 0.1 and 0.6 
by the HL-LHC searches.
The monotop signature will provide sensitivity up to $m_{\phi}\sim1600$~GeV
for $D_{\lambda,11}=1$ {and will probe the thermal freeze-out scenario for mediator masses above 1100~GeV. Between 800 and 1200~GeV the thermal relic assumption will be tested by the $tj$ analysis.}

\section{Conclusion and outlook}

Abundant production of final states with two quarks of different 
flavours accompanied by \etmiss from dark matter particles is a
well defined prediction of {flavoured DM models within the DMFV framework}.  
A particularly interesting case is when one of the two quarks 
is a top quark, which, with its distinctive decay signature, provides
an excellent experimental handle for searching for this model {at colliders}.

In the present study, we {addressed} three signatures with a single
top in the final state, accompanied by \etmiss and by no 
additional jet or a light jet or a $b$-jet.
For each of these signatures, we fully developed a search strategy 
at a 14~TeV LHC, taking into account all relevant standard model
backgrounds. The potential of the proposed new searches was
tested on four {phenomenologically viable} benchmark models with only two free parameters,
where all the remaining parameters of the model are fixed 
by low-energy or astrophysical constraints.
We compared the reach of the proposed signatures with 
the recast of existing flavour-conserving SUSY limits obtained
at the LHC with 13 TeV center-of-mass energy and an integrated luminosity
of 137~fb$^{-1}$. 

It turns out that for all of the considered benchmarks
the {final states} with a single top do increase the reach of the existing
analyses in the considered parameter space, and there are large 
regions in parameter space where the signal from several different 
analyses would be detectable, providing a handle on model discrimination.
A projection of the reach of the proposed analyses to the full LHC
luminosity and to the luminosity of the HL-LHC shows that for all of the
benchmark models {mediator masses of 1.6~TeV or above can be probed for a DM coupling parameter} $D_{\lambda,11}=1$,
{with the reach} decreasing for lower values of $D_{\lambda,11}$, where the final state 
with two top quarks and \etmiss will {eventually be} the most sensitive
channel. 

A monotop analysis optimised for these models along with searches 
for single top quarks accompanied by jets would {hence} provide the experimental
collaborations {with} a window to dark matter models incorporating flavour violation,
which have received only passing attention in the early LHC analyses.

\section*{Acknowledgements}

  The research of MB is supported by the DFG Collaborative Research Center TRR 257 ``Particle Physics Phenomenology after the Higgs Discovery''.

\bibliographystyle{JHEP}
\bibliography{dmfv}
\end{document}